\documentclass{article}

\def\*{{\bf ***}}

\def\a{\alpha}
\def\b{\beta}

\def\ga{\gamma}
\def\de{\delta}   
\def\eps{\varepsilon}
\def\phi{\varphi}
\def\la{\lambda}

\def\s{\sigma}

\def\om{\omega}

\def\F{{\cal F}}
\def\G{{\cal G}}

\def\J{{\bf J}}

\def\K{{\cal K}}

\def\R{{\bf R}}

\def\T{{\rm T}}

\def\Ga{\Gamma}
\def\De{\Delta}

\def\pa{\partial}

\def\d{{\rm d}}       

\def\o+{\oplus}

\def\ss{\subset}
\def\sse{\subseteq}

\def\<{\langle}
\def\>{\rangle}

\def\EOR{~ \hfill {$\odot$}}
\def\EOP{~ \hfill {$\diamondsuit$} \medskip }

\def\interno{\hskip 2pt \vbox{\hbox{\vbox to .18
truecm{\vfill\hbox to .25 truecm
{\hfill\hfill}\vfill}\vrule}\hrule}\hskip 2 pt}

\def\({\left(}
\def\){\right)}
\def\[{\left[}
\def\]{\right]}
\def\=#1{\bar #1}
\def\~#1{\widetilde #1}
\def\.#1{\dot #1}
\def\^#1{\widehat #1}
\def\"#1{\ddot #1}

\def\mapright#1{\smash{\mathop{\longrightarrow}\limits^{#1}}}
\def\mapdown#1{\Big\downarrow\rlap{$\vcenter{\hbox{$\scriptstyle#1$}}$}}

\def\ket#1{{| {#1} \rangle}}

\def\ref#1{\cite{#1}}

\begin{document}

\title{Asymptotic scaling symmetries for nonlinear PDEs}

\author{Giuseppe Gaeta\footnote{e-mail: {\tt gaeta@mat.unimi.it}} \\ 
{\it Dipartimento di Matematica, Universit\`a di Milano } \\
{\it via Saldini 50, I--20133 Milano (Italy)}
\\ 
Rosaria Mancinelli\footnote{e-mail:
{\it mancinelli@fis.uniroma3.it}; supported by {\it INFM}.} \\
{\it Dipartimento di Fisica, Universit\`a di Roma Tre} \\
{\it via della Vasca Navale 84, I--00146 Roma (Italy)}}

\maketitle

{\bf Summary.}
In some cases, solutions to nonlinear PDEs happen to be
asymptotically (for large $x$ and/or $t$) invariant under a group
$G$ which is not a symmetry of the equation. After recalling the
geometrical meaning of symmetries of differential equations -- and
solution-preserving maps -- we provide a precise definition of
asymptotic symmetries of PDEs; we deal in particular, for ease of
discussion and physical relevance, with scaling and translation
symmetries of scalar equations. We apply the general discussion to
a class of ``Richardson-like'' anomalous diffusion and
reaction-diffusion equations, whose solution are known by
numerical experiments to be asymptotically scale invariant; we
obtain an analytical explanation of the numerically observed
asymptotic scaling properties. We also apply our method to a
different class of anomalous diffusion equations, relevant in
optical lattices. The methods developed here can be applied to
more general equations, as clear by their geometrical
construction.

\section*{Introduction}

Symmetry methods for the study of differential equations were
first introduced by S. Lie; they are by now a widely known and
very effective set of tools to tackle nonlinear equations, both in
the sense of theoretical and geometrical study of their properties
and for obtaining exact solutions. Among the fastly growing
literature on these, we will here just refer to
\cite{Gae,Kra,Olv1,Ste}; see also the shorter introductions given
in \cite{CG,Ibr,Win}.

Once an algebra $\G_0$ of vector fields generating a Lie group $G_0$ of symmetries for a differential equation (or system of differential equations) $\Delta$ has been identified, there exist well defined methods to obtain $G_0$-invariant solutions to $\Delta$ via a reduced equation $\Delta_0$, sometimes denoted as $\De_0 = \De / G_0$.
These are also called {\it partially invariant solutions}, as in general $\G_0$ is only a part of the full symmetry algebra of $\De$.
This method is exposed in detail in most books on symmetry methods for differential equations, see e.g. \cite{Olv1} or \cite{Win}, and here we will just briefly recall it.

Other methods, generalizing or extending this approach, have been
developed by several researchers \cite{OlV}; for recent and quick
reviews see e.g. \cite{CiK,GTW}. Among these we mention in
particular the method of conditional symmetries and conditionally
invariant solutions \cite{ClK,Fus1,LeWin,PuS}; and an extension of
this, the method of partial symmetries\footnote{The reader should
be warned that the terms ``partial symmetry'' or ``partial
invariance'' are also used with a different meaning in the
literature, see e.g. \cite{OlV,Ovs}.} \cite{CGpar} (related in
turn to ``weak symmetries'' \cite{OlR}), which will be of use
later on in our discussion.

No matter how successful the symmetry approach, it has long been
known that there are equations which exhibit asymptotically stable
solutions $u_* (x,t)$ which are invariant under an algebra $\G_*$
which is {\bf not} a symmetry of the equation; moreover, solutions
which are asymptotic  to $u_* (x,t)$ are in general {\bf not}
invariant under $\G_*$, so that $\G_*$ is only an {\it asymptotic
symmetry} of these.\footnote{Among equations exhibiting such a
behaviour are a wide class of diffusion and reaction-diffusion
equations related to anomalous diffusion (studied numerically in
\cite{MVV2}, to which we refer for a discussion of their physical
significance), which we will study in this note as a nontrivial
example of application of our method; these will be called
``Richardson-like'' as they include the Richardson equation for
developed turbulence. Another class of equations with such a
behaviour -- also studied below -- describes the marginal Wigner
distribution in certain optical lattices \cite{Abe,Lut}, which
have recently attracted some attention due to the unexpected
appearance of Tsallis distribution as stationary states.}

Several authors have attempted a description of asymptotic invariance
(mainly, scaling and/or translational) properties of the
attracting asymptotic solution $u_* (x,t)$ mentioned above; it has
been generally agreed that such a description should be based on
(a suitable version of) {\it renormalization group} concepts.
We recall here the books by Barenblatt \cite{Bar,Bar2} and by Collet and Eckmann \cite{CE}, together with the works of Bricmont and
Kupiainen \cite{BK} and of Goldenfeld {\it et al.} \cite{Gold}. We
also quote the rather abstract approach proposed some time ago by
one of us \cite{Gae}, which will be developed here, focusing on the physically most relevant case of asymptotic scaling symmetries, into a concrete theory providing an explanation to observed asymptotic symmetry properties in the class of equations mentioned above, and possibly prediction of similar properties in other systems.\footnote{We stress that we focus on asymptotics for large
$t$, and in the cases of interest here this will correspond to
large $|x|$ as well. Asymptotic -- or ``approximate'' --
symmetries for small $|x|$ have been considered by a number of
authors, see e.g. \cite{BGI,CG,Fus2,Gio}.}

As mentioned above, in this note we will consider scalar PDEs (and
focus in particular on those of reaction-diffusion type associated
to anomalous diffusion; see e.g. \cite{CLV} for a recent review,
focusing on aspects of interest here), and accordingly we will
adopt suitably simplified notation and definitions; note however
that the methods developed here do generalize to the vector case.

\bigskip
The paper is organized as follows. In sect.1 we introduce in some
length the basic concepts of symmetry for differential equations;
and in sect.2 we recall the notions of conditional and partial
symmetry. We pass then to new material. In sect.3 we introduce our
definition of asymptotic symmetries (discussing directly the case
of second order scalar evolution equations), and in sect.4 we
extend this to conditional and partial asymptotic symmetries. In
sect.5 we discuss how these can be used to investigate the
asymptotic behaviour of an evolution PDE.

The rest of the paper is mainly devoted to a detailed
investigation of a model class of diffusion and reaction-diffusion
(RD) equations related to ``Richardson-like'' anomalous diffusion,
and their asymptotic behaviour, making use of the method proposed
here. In sect.6 we deal with the anomalous diffusion case, which
turns out to be rather simple. In sect. 7 we discuss the standard
FKPP equation from the point of view proposed in this paper,
discussing in particular the asymptotic symmetries of the equation
and of its asymptotic solutions. In sect.8 we consider
Richardson-like FKPP equations, recall the outcome of numerical
experiments on these, and determine their asymptotic
scaling-invariant solutions; this provides an explanation of
experimentally observed scaling properties. In sect.9 we consider
again Richardson-like FKPP equations, but look at more general
asymptotic symmetries, focusing on large $t$ behaviour near a
propagating front. It turns out this analysis provides again a
direct explanation of the asymptotic solutions observed in
numerical experiments.

In section (10) we consider a different anomalous diffusion
equation, i.e. the one describing the marginal Wigner distribution
in momentum space of an atom in an optical lattice \cite{Lut};
attention has been recently called to this not only for its
relevance in applications, but also because -- somewhat
surprisingly -- its asymptotic, time-invariant, solutions are in a
certain regime scale-free and correspond to a Tsallis statistics
\cite{Tsa}. We show that this behaviour can be explained on the
basis of the asymptotic symmetry properties (under generalized
scalings) of the equation.

The final sect.11 is devoted to summarizing and discussing our
findings, and presenting some final remarks.

\bigskip\noindent
{\bf Acknowledgements.} We thank D.Levi for useful discussions.
The work of GG was supported in part by {\it GNFM--INdAM} under
the program {``Simmetrie e tecniche di riduzione per equazioni
differenziali''}; the work of RM was supported by {\it I.N.F.M.
(Istituto Nazionale di Fisica della Materia)}.

\section{Symmetries of differential equations}

In this section we recall the main concepts and definitions for
symmetries of differential equations (see e.g.
\cite{Gae,Olv1,Ste,Win} for general treatments of these). As in
this note we will only consider scalar PDEs, we will just
specialize our formulas to this case.

\subsection{General notions and properties}

We consider an equation of order $n$ for $u = u (x)\in \R$, with
$x \in \R^m$. We denote by $B = \R^{m}$ the space of independent
variables, by $U = \R$ the space of dependent variables, and by $M
= B \times U$ the total space of independent and dependent
variables.

Note that $M$ has a natural structure of fiber bundle over $B$, i.e. $(M,\pi,B)$ is a bundle with fiber $\pi^{-1} b = U$. A function $u = f(x)$ is then represented by a section $\ga_f$ of this bundle.

A vector field in $M$, written in coordinates as
$$ X \ = \ \xi^i (x,u) \, \frac{\pa}{\pa x^i} \ + \ \phi (x,u) \,  \frac{\pa}{\pa u} \ , \eqno(1.1) $$
will generate a one-parameter group of transformations in $M$; at the infinitesimal level these are described by
$ x^i \to \~x^i = x^i + \eps \xi^i (x,u)$, $u \to \~u = u + \eps \phi (x,u)$.

Thus, for $\eps$ sufficiently small, the function $u = f(x,t)$ is mapped to a new function $u = \~f (x,t)$; one obtains with standard computations that this new function is given by
$$ \~f (x) \ = \ f (x) \ + \ \eps \ \[ \phi (x,u) - \({\pa f (x) \over \pa x^i } \) \cdot  \xi^i (x,u) \]_{u=f(x)} \ + \ o (\eps) \eqno(1.2) $$
where $\pa_i := \pa / \pa x^i$; functions depending on $u$ should be computed in $u = f(x)$.

We introduce some standard notations: first of all, we denote by $D_i$ the total derivative with respect to $x^i$, i.e.
$ D_i := (\pa / \pa x^i) + u_i (\pa / \pa u) + u_{ij} (\pa / \pa u_j)   + ...$.
We will use multiindices $\J = (j_1,...,j_m)$; the order of $\J$ is $|\J| = j_1 + ... + j_m$. With these, we write $D_\J := (D_i)^{j_1} ... (D_m)^{j_m}$, and $u_\J = D_{\J} u$; we also write $u_{\J,i} := D_i u_\J$.
Finally, the space of $(x,u)$ and $x$-derivatives of $u$ up to order $k$ is said to be the {\it jet space} of order $k$ for $M$, and denoted by $J^{(k)} M$; it is convenient to also use the formal limit $k \to \infty$, in which case we write $J^{(*)} M$.

By differentiating (1.2), we get formulas for the transformation of the partial derivatives of any order of $u$ with respect to $x$ and $t$ under the action of $X$.

The action of the vector field $X = \xi^i \pa_i + \phi \pa_u$ on the jet space $J^{(*)} M$ is described by the {\bf prolongation} of $X$, i.e. by the vector field
$$ X^* \ := \ X \ + \ \sum_\J \Psi_\J {\pa \over \pa u_\J} \eqno(1.3) $$
where the coefficients are given by the {\bf prolongation formula}
$$ \Psi^a_\J \ = \ D_\J (\phi^a - u^a_k \xi^k ) + u^a_{\J,k} \xi^k \ ; \eqno(1.4) $$
this reads also, in recursive form and with $\Psi^a_0 := \phi^a$,
$$ \Psi^a_{\J,k} \ = \ D_k \Psi^a_\J - u_{\J,i} D_k \xi^i \ . \eqno(1.5) $$

The differential equation $\De$ of order $n$ will be written as
$\Phi (x,u,...,u^{(n)} ) = 0 $ for some function $\Phi :
J^{(n)} M \to \R$.
We can then apply the vector field $X^{(n)}$ to $\Phi$ and
consider the result of this\footnote{Note that if we apply
$X^{(*)}$ to $\Phi$, only terms corresponding to $|\J| \le n$ in
(1.3) will actually matter: we can as well consider the (formal)
infinite-order prolongation.} once the constraint $\Phi=0$, i.e.
the equation $\De$ itself, has been taken into account.

If $ \Phi = u_t - F (x,t,u,u_x,u_{xx})$, as in our
applications below, taking $\De$ into account consists simply in
writing $F (x,t,u,u_x,u_{xx})$ for $u_t$ (this also extends to
differential consequences: e.g., $u_{xt} = D_x F$).

We say that $X$ is a {\it symmetry} of the equation $\De$ given by
$\Phi=0$ if\footnote{$X$ is a {\it strong symmetry} of $\De$ if
(1.6) holds without the restriction to $\De$ (i.e. to $\Phi=0$).
Any strong symmetry is also, of course, an ordinary symmetry; see
\cite{CDW} for details of the relation between ordinary and strong
symmetries.}
$$ \[ X^{(n)} (\Phi) \]_{\Phi=0} \ = \ 0 \ . \eqno(1.6) $$

Condition (1.6) guarantees that $X$ maps solutions to $\De$ into
(generally, different) solutions to $\De$; conversely, if $X$ maps
solutions into solutions, then (1.6) is necessarily satisfied.
Actually, the qualitative meaning of {\it symmetries of a
differential equation} is precisely that of transformations
mapping any solution to $\De$ into a solution
\cite{GaeK,Olv1,Ste,Win}.

These concepts can also be stated in an equivalent but more geometric way, more convenient for our discussion, as follows. The equation $\De$ of order $n$ identifies a {\it solution manifold} $S$ in $J^{(n)} M$; a function $u = f(x)$ is a solution to $\De$ if and only if the prolongation $\ga_f^{(n)}$ of the section $\ga_f$ satisfies $\ga_f^{(n)} \ss S$.
The vector field $X$ is a symmetry for $\De$ if its prolongation $X^{(n)}$ to $J^{(n)} M$ is tangent to $S$, i.e.
$$ X^{(n)} \ : \ S \, \to \, \T S \ . \eqno(1.6') $$

We stress that in this note we are mainly interested in two kinds of symmetries: {\it scalings} and {\it shifts} (or translations). We recall that the shift $x^i \to x^i + \eps s_i $, $u \to u + \eps s_0$   (with $s_j \in \R$) is generated by $X = s_i \pa_i + s_0 \pa_u$, and the scaling $x^i \to \la^{k_i} x^i$, $u \to \la^{k_0} u$ (with $k_i \in \R$) is generated by $X = k_i x \pa_i + k_0 u \pa_u$.

\subsection{Symmetry reduction for PDEs}

Let $\De$ be a PDE for $u = u(x)$, $x \in \R^p$, admitting a vector field $X$ as symmetry; in order to determine $X$-invariant solutions to $\De$ one proceeds in the following way. (This is standard material, recalled here to fix notation; for more details, see e.g. the discussion in chap.3 of \cite{Olv1}.)

We write $\De$ in the form $F (x,u^{(k)}) = 0 $, with $F$ a smooth scalar function, $F : J^{(k)} M \to \R$, and denote by $Y$ the $k$-th prolongation of $X$. For reaction-diffusion equations, $p=2$ and $k=2$.

First of all we pass to symmetry-adapted coordinates in $M$. In practice, we have to determine a set of $p$ independent invariants for $X$ in $M$, which we will denote as $(y^1,...,y^{p-1};v)$: these will be our $X$-invariant coordinates, and essentially identify the $G$-orbits, while the remaining coordinate $\s$ will be acted upon by $G$. In other words, $G$-orbits will correspond to fixed value of the $(y,v)$ coordinates, with $\s$ taking values in a certain subset of the real line.

The invariants will be given by some functions of the $x$ and $u$, i.e.
$$ y^i = \eta^i (x,u) \ (i=1,...,p-1) \ , \ \ v = \zeta (x,u) \ , \eqno(1.7) $$
with $\s = \s (x,u)$ as well. Assume (this is always the case for $X$ a scaling vector field) that we can invert the above for $x$ and $u$ as functions of $(y,v;\s)$:
$$ x^i = \chi^i (y,v;\s) \ (i=1,...,p) \ , \ u = \b (y,v;\s) \ . \eqno(1.8) $$

If now we decide to see the $(y;\s)$ as independent variables and the $v$ as the dependent one, we can use the chain rule to express $x$-derivatives of $u$ in terms of the  $\s$ and $y$-derivatives of $v$. Using these, we can finally write $\De$ in terms of the $(\s,y;v)$ coordinates, i.e. in the form $\^\Phi (\s , y , w^{(n)})$. As $X$ is a symmetry of $\De$, it follows that the function $\^\Phi$, when subject to the side condition $\pa v / \pa \s = 0$, is independent of $\s$.

In fact, $\pa v / \pa \s = 0$ expresses the fact that the
solutions $v = v(y,\s )$ are required to be invariant under $X$: the equation obtained in this way represents the restriction of $\De$ to the space of $G$-invariant functions, and is therefore also denoted as $\De /G$.

Suppose we are able to determine some solution $v = \^f (y)$ to the reduced equation; we can write this in terms of the $(x,u)$ coordinates as
$$ \zeta (x,u) \ = \ \^f [ \eta (x,u) ] \eqno(1.9) $$
which yields implicitly $u = f (x)$, the corresponding $G$-invariant solution to the equation $\De$ in the original coordinates.

\medskip\noindent
{\bf Remark 1.} Given any vector field $X$ we can consider its {\it characteristic} $Q := \phi - u_k \xi^k$ and its {\it evolutionary representative} $ X_Q := Q  (\pa / \pa u)$.
By (1.4) or (1.5), the prolongation of the latter will simply be
$ X_Q^* = X_Q + \sum_\J (D_\J Q^a) ( \pa / \pa u^a_\J) $,
and standard computations \cite{Olv1} show that
$ X^* = X_Q^* + \xi^k D_k $. These formulas give a very convenient way of computing prolongations.

The reduction procedure described above can also be described in a slightly different way using evolutionary representatives: if we look for $X$-invariant solution $u=f(x)$ to $\De$, we determine the characteristic $Q = \phi - u_i \xi^i$ of the vector field $X$, and supplement $\De$ with the equations $D_J Q = 0$ with $|J|=0,...,k-1$. The equations $Q = 0$ require that the evolutionary representative $X_Q = Q (\pa / \pa u)$ of $X$ vanish on $\ga_f$, i.e. that $u$ is $X$-invariant, and all the equations with $|J| >0$ are just differential consequences of this. The $X$-invariant solutions to $\De$ are in one to one correspondence with the solutions to the system $\De_X := \{ \De ; D_J Q = 0 \}$. See e.g. \cite{Win} for details, and for how this approach is used in a more general
context. \EOR

\medskip\noindent
{\bf Remark 2.} The standard method discussed here applies under a nondegeneracy (transversality) condition, guaranteeing a certain Jacobian admits an inverse. When this is not the case the treatment should go through the approach developed by Anderson, Fels and Torre \cite{AFT}; in case of partial transversality see also \cite{GTW}. Also, this method is justified only if the (possibly, only local) one-parameter group generated by $X$ has regular action in $M$. Both these conditions are satisfied for scaling vector field other than the trivial $X_0 = u \pa_u$ (which is a symmetry only for linear equations), and translations other that $X = \pa_u$ (which is a symmetry for equations of Hamilton-Jacobi type). \EOR

\subsection{Solution-preserving maps}

Consider a PDE $\De$, say of the form
$$ \De \ \equiv \ u_t \, - \, F (x,t,u,u_x,u_{xx} ) \ = \ 0 \ , \eqno(1.10) $$
with $(x,t,u) \in M = \R^3$, and $M$ a bundle $(M,\pi,B)$ over $B = \R^2$. Let us now consider another manifold $\^M = \R^3$ which is also  a bundle $(\^M , \^\pi , \^B )$ over $\^B = \R^2$, and a ${\cal C}^n$ map $F : M \to \^M$ mapping $(x,t,u) \in M$ into $\xi,\tau, w) \in \^M$, with of course  $\xi = \xi (x,t,u)$, $\tau = \tau (x,t,u)$, $w = w (x,t,u) )$. This can be lifted to a map $F^{(n)} : J^{(n)} M \to J^{(n)} \^M$, once we identify $w$ as the dependent variable in $\^M$.
In this way, $(u_t,u_x;u_{tt},u_{xt},u_{xx} )$ are also mapped to expressions involving $(w_\tau , w_\xi ; w_{\tau \tau}, w_{\xi \tau} , w_{\xi \xi} )$.

The geometrical meaning of this is that the solution manifold $S \ss J^{(n)} M$ of $\De$ is mapped by $F^{(n)}$ into the solution manifold $\^S \ss J^{(n)} \^M$ of $\^\De$.

If the map is projectable \cite{GaeK}, i.e. such that $\tau = \tau
(t)$, $\xi = \xi (x,t)$, then (1.10) is mapped into an equation of
the same type:
$$ \^\De \ \equiv \ w_\tau \, - \, G (\xi,\tau,w,w_\xi,w_{\xi \xi} ) \ = \ 0 \ . \eqno(1.11) $$

Recall now that the function $u = f(x,t)$ corresponds with the section $\ga_f = \{ (x,t, f(x,t) ) \}$ in $(M,\pi,B)$.
If $F$ is projectable, we are guaranteed that the manifold $\ga_f \ss M$ is mapped into a manifold $\ga_g \ss \^M$ which is a section for $(\^M , \^\pi , \^B )$. In other words, $F (\ga_f ) = \ga_g = \{ (\xi , \tau , g (\xi , \tau ) ) \} \ss \^M$. This obviously extends to prolongations, i.e. $F^{(n)} : \ga_f^{(n)} \to \ga_g^{(n)}$.
This guarantees, in particular, that solutions $u = f (x,t)$ to $\De$ are mapped into solutions $w = g (\xi , \tau)$ to $\^\De$.
We say therefore that $F$ is a {\it solution preserving} map \cite{Wal}.

If $F$ is invertible (with a ${\cal C}^n$ inverse), we can repeat these considerations for $\Psi = F^{-1}$. In this case, therefore, the equations $\De$  and $\^\De$ are {\it equivalent}, in that there is a ${\cal C}^n$ isomorphism between solutions to $\De$ and solutions to $\^\De$.

\section{Conditional and partial symmetries}

As mentioned above, $X$ is a symmetry if it maps any solution to a
(generally, different) solution. However, there will be vector
fields which map some solutions into solutions, and some other
solutions into functions which are not a solution to $\De$. In
this case we speak of partial symmetries.

That is, $X$ is a {\it partial symmetry} for $\De$ if there is a
nonempty set ${\cal S}_X$ of solutions to $\De$ which is mapped to
itself by $X$. Note that ${\cal S}_X$ could be made of a single
solution, and more generally that there could be solutions which
are left invariant by $X$. In this case, we say that $X$ is a {\it
conditional symmetry} for $\De$. Obviously, any symmetry is also a
conditional symmetry, and any conditional symmetry is also a
partial symmetry.\footnote{Partial symmetries (as defined here)
were introduced and characterized in \cite{CGpar}. The origin of
conditional symmetries could maybe be traced back to Felix Klein
\cite{Gae}; for their modern theory, see \cite{ClK,Fus1,LeWin}
(here we adopt the point of view of \cite{LeWin}, see also
\cite{Win}). For the relation of conditional symmetries to the
``direct method'' of Bluman and Cole \cite{BlC}, see \cite{PuS}.}

It is clear from (1.2) that $u = f(x)$ is invariant under $X$ if and only if
$$ \De_X \ := \ \phi [x,f(x)] - u_i \xi^i [x,f(x)] = 0 \ ; \eqno(2.1) $$
thus $X$-invariant solutions to $\De$ are solutions to the system $\^\De$ of differential equations made of $\De$ and $\De_X$. In other words, $X$ is a conditional symmetry of $\De$ if and only if it is a symmetry of $\^\De$.

Partial symmetries can be seen in a similar way. As discussed in \cite{CGpar}, a function $f$ in the globally invariant set of solutions to $\De : F = 0$,  $ f \in {\cal S}_X$ in the notation used above,  will be a solution to a system
$$ \cases{
F^{(0)} = 0 & \cr
F^{(1)} = 0 & \cr
...... & \cr
F^{(p)} = 0 & \cr} \eqno(2.2) $$
where $F^{(0)} \equiv F$, and $F^{(k+1)} := X^* [ F^{(k)} ]$. The integer $p$, i.e. the order of the system, is determined as the lowest order such that $F^{(p)}$ vanishes identically on solutions to the previous equations. Note also that each equation $F^{(k)}=0$ can, and should, be simplified by taking into account the previous equations.

Needless to say, if $\De$ and $\^\De$ are related by a
solution-preserving map $\Phi$, this entails a relation between
their respective conditional and/or partial symmetries.

For a recent discussion on partial symmetries and related notions,
see \cite{CiK}; for examples, see \cite{CGpar}.

\section{Asymptotic scaling symmetries for evolution PDEs of second order}

As suggested by their names, {\it asymptotic symmetries} of a differential equation $\De$ are vector fields $X$ which, albeit in general not symmetries of $\De$, satisfy $X^{(n)} : S_\De \to \T S_\De$ asymptotically (see below for the precise sense of this). Any exact symmetry is also a (trivial) asymptotic symmetry.

Asymptotic symmetries were introduced and discussed in quite a general framework in \cite{Gae}, see also \cite{CG}. In this section we introduce them in a simplified scheme, i.e. limiting our considerations to the class of equations of interest here (evolution PDEs, in particular of reaction-diffusion type) and to the  transformations we are interested in (scalings and shifts).

\subsection{Asymptotic symmetries for equations of RD type}

Thus, we discuss scalar equations $\De$ for a real function of two variables, $u (x,t)$. In the notation of sect.1 this means $B = \R^2$, $M = \R^3$. We write the equation $\De$ in the form $u_t - F (x,t,u,u_x,u_{xx}) = 0$.

We denote by $\F$ the space of maps $F : (x,t,u,u_x,u_{xx}) \to \R$ which are polynomial in $(u,u_x,u_{xx})$ and rational in $(x,t)$. Note that this space, which corresponds to the space of equations in the form we are considering, is invariant under scaling transformations and under translations.

Vector fields will be written as $X = \tau \pa_t + \xi \pa_x + \phi \pa_u$, and we mainly restrict to scaling ones, i.e.
$$ X \ = \ (a t) \, \pa_t \ + \ (b x) \, \pa_x \ + \ (c u ) \, \pa_u \ \ . \eqno(3.1)  $$
This generates the one-parameter group of scaling transformations
$$ t \to \la^a t \ , \ x \to \la^b x \ , \ u \to \la^c u \ . \eqno(3.2) $$

The second prolongation of $X$, which we denote for ease of writing as $Y \equiv X^{(2)}$, is again a scaling vector field,
$$ \begin{array}{rl}
Y \ =& X \ + \ [(c-a) u_t] \pa_{u_t} + [(c-b) u_x] \pa_{u_x} \\
& + [(c-2a) u_{tt} ] \pa_{u_{tt}} + [(c-a-b) u_{xt}] \pa_{u_{xt}} + [(c-2b) u_{xx}] \pa_{u_{xx}} \ , \end{array} \eqno(3.3) $$
and generates the scaling group
$$ \begin{array}{l}
t \to \la^a t \ , \ x \to \la^b x \ , \ u \to \la^c u \ , \\
u_t \to \la^{c-a} u_t \ , \ u_x \to \la^{c-b} u_x \ , \\
u_{tt} \to \la^{c - 2 a} u_{tt} \ , \ u_{xt} \to \la^{c-a-b} u_{xt} \ , \ u_{xx} \to \la^{c-2b} u_xx \ . \end{array} \eqno(3.4) $$

Under the action of $X$, the function $u = f (x,t)$ is transformed  into $u = \~f (x,t)$ with, see (1.2),
$$ \~f (x,t) \ = \ f (x,t) \ + \ \eps \, \[ u - a t u_t - b x u_x \] \ + \ o (\eps ) \ . \eqno(3.5) $$
Thus $X$ induces a flow $\^X$ in the space of functions $u = f(x,t)$, or more precisely in the space of sections of the bundle $(M,\pi,B)$.
Note that $\^X$ corresponds to the evolutionary representative $X_Q$ of $X$, see sect.1.

We write the flow issued by $f_0 (x,t)$ as
$ f_\la (x,t) = \exp [{\la \^X}]$, with $d f_\la / d \la = \^X [ f_\la ]$. We say that $f_0 (x,t)$ is $X$-invariant if it is a fixed point for the flow of $\^X$, i.e. if $\^X [ f_0 ] = 0$.

It is also possible that $f_0$ is not invariant under $X$, but the flow $f_\la$ is asymptotic to an invariant function $f_*$, i.e.
$$ \lim_{\la \to \infty} | f_\la (x,t) - f_* (x,t) | = 0 \ \ , \ \ \^X [ f_* ] = 0 \ . \eqno(3.6) $$
When (3.6) is satisfied, we say that $f$ is asymptotically $X$-invariant under the flow of $\^X$.

\subsection{Symmetries in the space $\F$}

Let us now consider an equation $\De : \Phi (x,t,u^{(2)} )=0$ written as
$$ \De \ : \ \ u_t \, - \, F_0 (x,t,u , u_x , u_{xx}  ) \ = \ 0 \ .  \eqno(3.7) $$
Such an equation is thus identified by the function $F_0$, and we assume $F_0 \in \F$.

The action of $X$ on the space of such equations is described by the second prolongation $Y=X^{(2)}$. In order to consider the flow induced by $Y$ on the space of equations of the form (3.7), i.e. on $\F$, we write
$$ \De_\la \ = \ e^{\la Y} \De_0 \ := \ \s (\la) \, \[ u_t - F_\la (x,u,u_x,u_{xx} ) \] \ ;  \eqno(3.8) $$
by construction, this satisfies
$$ {d \De_\la \over d \la } \ = \ Y ( \De_\la ) \ . \eqno(3.9) $$

With $\De$ as in (3.7),
$$ Y [ \De ] \ = \ \la^{c-a} u_t - \[ \la^b x {\pa F \over \pa x} + \la^c u {\pa F \over \pa u} + \la^{c-b} u_x {\pa F \over \pa u_x} + \la^{c-2b} u_{xx} {\pa F \over \pa u_{xx} } \] \ . \eqno(3.10) $$

We wish to identify $\De_\la$ with the corresponding $F_\la$, see again (3.7). Hence we also write, comparing (3.7) and (3.10),
$$ {d F \over d \la} \ = \ \la^{a+b-c} x {\pa F \over \pa x} + \la^{-a} u {\pa F \over \pa u} + \la^{a-b} u_x {\pa F \over \pa u_x} + \la^{a-2b} u_{xx} {\pa F \over \pa u_{xx} } \ . \eqno(3.11) $$

In this way, $X$ induces (via $Y$) a vector field $W$ in the space $\F$; we rewrite (3.8) as $d F / d \la = W (F)$.

As recalled above, $X$ is a symmetry of $\De$ if and only if $Y : S \to \T S$ (recall $Y = X^{(2)}$). This condition is now rephrased in terms of $\F$ by saying that $X$ is a symmetry of $\De_0$ given by (3.7) if and only if $F_0$ is a fixed point for the flow of $W$.

\subsection{Asymptotic symmetries in the space $\F$}

Suppose now that $X$ is not a symmetry of $\De_0$ -- $F_0$ is not a fixed point for the flow of $W$ in $\F$ -- but that the flow issued by $F_0$ under $W$ satisfies
$$ \lim_{\la\to \infty} | F_\la - F_* | \ = \ 0 \ \ , \ W (F_* ) = 0 \ . \eqno(3.12) $$
In this case we say that $X$ is an {\it asymptotic symmetry} for $F_0$, i.e. for the equation $\De_0$.

\medskip\noindent
{\bf Remark 3.} The procedure given here to define asymptotic symmetries can be reinterpreted in a slightly different way: that is, we combine the action of $W \approx Y = X^{(2)}$ with a rescaling in the equation, so to keep this in the form $u_t - F = 0 $. \EOR
\medskip

By construction, and by the invertibility of the scaling transformation (3.4) for $\la$ finite, if $u = f (x,t)$ is a solution to $\De_0$, then $u = f_\la (x,t)$ will be a solution to $\De_\la$, and conversely if $u = f_\la (x,t)$ is a solution to $\De_\la$, then $u = f (x,t)$ is a solution to $\De_0$.

This is represented in the following diagram, where $S_\la \ss J^{(2)} M$ is the solution manifold for $\De_\la$:
$$ \matrix{
u_0 = f_0 (x,t) & \approx & \ga_0 & \ss & S_0 & \approx & \De_0 \cr
  & & \mapdown{e^{\la Y}} & & \mapdown{e^{\la W}} & & \mapdown{e^{\la W}} \cr
u_\la = f_\la (x,t) & \approx & \ga_\la & \ss & S_\la & \approx & \De_\la \cr} \eqno(3.13) $$

In the limit $\la \to \infty$ the invertibility of (3.4) fails; we can nevertheless still say that solutions to the original equation flow into solution to the asymptotic equation, provided both limits $f_*$ and $\De_*$ exist.

If the limit $\De_*$ exists, it captures the behaviour of $\De$ for large $\la$, i.e. for large (or small, depending on the sign of $a$ and $b$) $t$ and $|x|$.

\subsection{Solution-preserving maps and asymptotic symmetries}

If there is an equation $\^\De$ whose asymptotic behaviour is well understood, and such that there exists a solution-preserving map $\Phi : \De \to \^\De$, we can of course study the asymptotic behaviour of solutions $u(x,t)$ to $\De$ by means of the asymptotic behaviour of solutions $w (\xi , \tau )$ to $\^\De$. This approach is schematized in the diagrams below:
$$ \matrix{\De_0 & \mapright{\Phi^{(n)} } & \^\De_0 \cr
\mapdown{e^{\la W}} & & \mapdown{e^{\la \^W}} \cr
\De_\la & \mapright{\Phi^{(n)} } & \^\De_\la \cr} \ \ \ \ \ \ \ \
\matrix{u_0 & \mapright{\Phi} & w_0 \cr
\mapdown{e^{\la X}} & & \mapdown{e^{\la X}} \cr
u_\la & \mapright{\Phi} & w_\la \cr} \ \eqno(3.14)  $$
Here $\^W := \Phi^{(n)}_* (W)$ is the push-forward \cite{Nak,NaS} of the map $\Phi^{(n)}$, i.e. of the lift of $\Phi : M_1 \to M_2$ to a map $\Phi^{(n)} : J^{(n)} M_1 \to J^{(n)} M_2$, to the tangent map
$\Phi^{(n)}_* : \T [J^{(n)} M_1] \to \T [J^{(n)} M_2]$.

If $\Phi$ is a solution-preserving map, these diagrams commute; if $\Phi$ has a smooth inverse $\Psi$, we can study the flow of $W$ using $$ e^{\la W} \ = \ [ \Phi^{(n)} ]^{-1} \, \circ \, e^{\la \^W} \, \circ \, \Phi^{(n)} \ . \eqno(3.15) $$
Note that (3.15) remains valid also in the limit $\la \to \infty$, i.e. for the asymptotic regime.

Needless to say, this approach is particularly convenient when the asymptotic behaviour of the solutions $w (\xi , \tau) $ to $\^\De_0$ -- i.e. the behaviour of solutions to $\^\De_*$ -- is simple and/or in some way universal (even better if $\^\De_0$ is a fixed point under $\^W$).

Denoting as $w_* (\xi , \tau)$ the limit expression for the solutions to $\^\De_0$, the asymptotic behaviour of the solution $u (x,t)$ to $\De_0$ will be given by
$$ u_* (x,t) \ = \ \Phi^{-1} \, \[ w_* (\xi , \tau ) \] \ . \eqno(3.16) $$

More precisely, considering the sections $\ga_f$ and $\ga_g$ corresponding to $u = f (x,t)$ and $w = g (\xi , \tau)$ (see sect.1), and going asymptotically into $\ga_f^*$ and $\ga_g^*$ respectively, we have $\ga_f^* = \Phi^{-1} (\ga_g^* )$.

\section{Conditional and partial asymptotic symmetries}

It is also possible to consider asymptotic versions of conditional
and partial symmetries. These are introduced in accordance with
our general scheme.

That is, we could have the case where a function $u = f_0 (x,t)$
(more precisely, the corresponding section $\ga_0 \ss M$) flows
under $e^{\la X}$ to a fixed point $u = f_* (x,t)$ (more
precisely, a $X$-invariant section $\ga_* \ss M$) albeit $\De_0$
does {\it not} flow to a fixed point\footnote{In the general
framework \cite{Gae} the situation would be more complex, as the
invariance vector field $X$ would not in general be the vector
field applied to extract the asymptotic behaviour. Here, however,
we do not need to worry about such a general setting.}.

In such a situation, the solution manifold $S_\la$ does not go to a limit manifold, but there is a submanifold $S_\la^X$ of this, with $\ga_\la \sse S_\la^X \ss S_\la$, which flows to a fixed limit submanifold $S_*^X$, with $\ga_* \sse S_*^X$.

In this case we say that $X$ is a {\bf conditional asymptotic
symmetry} for $\De$. Examples of this phenomenon will be discussed
in sect.8 below, where we deal with generalized FKPP equations.

The same construction, with suitable and rather obvious modifications, applies for what concerns partial symmetries: suppose that $\De \equiv \De^{(0)}$ does not flow to a fixed point under the vector field $W$ induced by $X$ in $\F$. Consider the equations
$$ \De^{(1)} := Y [ \De^{(0)} ] \ , \ ... \ , \ \De^{(r)} := Y [ \De^{(r-1)} ] $$
up to an $r$ -- if it exists -- such that $\De^{(r)}$ does admit a fixed point $\De^{(r)}_*$ under the $W$ flow, while the $\De^{(k)}$ with $k < r$ do not. Then the manifold
$$ S_\la^{(0)} \cap ... \cap S_\la^{(r)} \ := \ {\cal S}_\la $$
(with $S_\la^{(k)} \ss J^{(n)} M$ the solution manifold for $\De_\la^{(k)}$) flows to a limit submanifold ${\cal S}_*$, and solutions $u = f_0 (x,t)$ to the system $\De^{(k)}$ ($k=0,...,r$) flow to functions $u = f_* (x,t)$ such that the prolongation $\ga_*^{(n)} \ss J^{(n)} M$ of the corresponding section $\ga_{f_*} = \{ (x,t,f_* (x,t)) \}$ lie in ${\cal S}_*$.
In this case we say that $X$ is a {\bf partial asymptotic symmetry} for $\De$.

\medskip\noindent
{\bf Remark 4.} If $\De$ and $\^\De$ are related by a solution-preserving map $\Phi$, this also relates their conditional and partial symmetries. In particular, if $\Phi : \De \to \^\De$ is invertible and $\^\De$ admits $X$ as a conditional symmetry, then $\De$ admits $\Phi^{-1}_* (X)$ as a conditional symmetry, with $\Phi^{-1}_*$ the push-forward of $\Phi^{-1}$. This will be of use below. \EOR

\section{Asymptotic symmetries as a tool to test asymptotic behaviour}

In many physically relevant cases, one has to study nonlinear PDEs
which are not amenable to an exact treatment, or at least for
which such a treatment is not known, and for which a numerical
study shows an asymptotic behaviour which appears to be well
described by some kind of invariance. Usually, the latter
corresponds to a scale invariance (self-similar solutions), or a
translation invariance (travelling waves), or a combination of
both of these.

The discussion conducted so far can be implemented into a
procedure allowing on the one hand to test if the observed
asymptotic behaviour is a characteristic of the equation (rather
than an artifact of the numerical experiments conducted on it),
and on the other hand to formulate simpler equations extracting
the asymptotic behaviour.

We will now describe the procedure in operational terms; as a
(rather simple, but relevant) example to illustrate our procedure
we will consider here the heat equation, while in later sections
we apply the procedure on equations associated to anomalous
diffusion.

We will denote by $X$ the vector field describing the observed invariance. For the sake of concreteness, we consider a second order  equation for $u = u(x,t)$ of the form $ u_t = F (x,t,u,u_x,u_{xx})$.
We denote, as usual, by $M$ the space $\R^3$ of independent and dependent variables, i.e. $M = \{ (x,t;u) \}$.

\begin{itemize}

\item {\bf Step 1.} Pass to symmetry-adapted coordinates in $M$,
i.e. coordinates $(\s , y ; v)$ such that $X(y) = X(v) = 0$; thus
in these coordinates $X = f(\s,y,v) (\pa / \pa \s )$. \footnote{We
stress that we are not requiring $f=1$; actually when we deal with
scaling symmetries it is appropriate to require $f (\s , y , v) =
\s$.}

\item {\bf Step 2.} Identify $v$ as the new dependent variable,
i.e. $v = v (\s , y)$. This allows to write $x$ and $t$
derivatives of $u$ as $\s$ and $y$ derivatives of $v$, hence to
write the differential equation $\De (x , t;u^{(2)} ) $ as $\^\De
(\s,y;v^{(2)} )$.

\item {\bf Step 3.} Reduce the equation $\^\De (\s,y;v^{(2)} )$ to
the space of $X$-invariant functions, i.e. to $v(\s,y)$ satisfying
$v_\s = 0$. For $X$ an exact symmetry, the reduced equation
$\^\De_X$ will not depend on $\s$ at all. For $X$ a conditional or
partial symmetry, $\s$ will still appear parametrically in the
reduced equation.

\item {\bf Step 4.} Study the asymptotic behaviour of the
solutions to the reduced equation $\^\De_X$ for $\s \to \infty$.

\item {\bf Step 5.} Go back to the original variables.

\end{itemize}

\subsection*{An elementary example: the heat equation}

Let us illustrate our procedure by applying it on the heat equation $u_t = u_{xx}$. Its asymptotic solutions are of the form
$$ u(x,t) \ = \ {A \over \sqrt{2 t} } \ \exp \[ - {4 x^2 \over t} \] \eqno(5.1) $$
and are invariant under the scaling vector field
$$ X \ = \ x \pa_x \, + \, 2 t \pa_t \, - \, u \pa_u \ , \eqno(5.2) $$
which is also an exact symmetry of the equation.

The symmetry-adapted coordinates (and the old ones in terms of these) are
$$ \begin{array}{lll}
\s = t \ ,& \ y = x^2 /t \ ,& \ v = x u \ ; \\
t = \s \ ,& \ x = \sqrt{\s y} \ , & u = v/\s \ . \end{array} \eqno(5.3) $$
Using these, the heat equation reads
$$ \s \, v_\s \ = \ 4 \, y \, v_{yy} \ + \ (2 + y) \, v_y \ + \ v \ . \eqno(5.4) $$
Imposing $v_\s = 0$ we obtain an equation with $\s$ disappearing completely. Needless to say, the equation obtained in this way has solutions
$$ v (y) \ = \ \widehat{A} \, \sqrt{y} \, \exp [- y/4] \ , \eqno(5.5) $$
which when mapped back to the original coordinates produce the gaussian (5.1).

Note also that (5.4) necessarily requires that for $\s \to \infty$, the function $v = v(\s,y)$ satisfies $v_\s = 0$: thus, the full asymptotic behaviour of (5.4) is captured by (5.5).

\section{Asymptotic symmetry of Richardson-like anomalous diffusion equations}

In this section we apply our procedure to anomalous diffusion (or
Fokker-Planck) equations
$$ u_t \ = \ \^L [u] \eqno(6.1) $$
with $\^L$ the linear operator
$$ \^L [u] \ := \ {x^{1 - \a/2} \over t^{1 - \nu \a} } \ {\pa \over \pa x} \[ x^{1 - \a/2} u_x \] \ . \eqno(6.2) $$

These will be called {\it "Richardson-like"} (in the following,
RL) since for $\a = 2/3$ and $\nu = 3/2$ we get the Richardson
equation
$$ u_t \ = \ {x^{2/3} } \ {\pa \over \pa x} \[ x^{2/3} u_x \] $$
describing the evolution of the distance between two particles in
developed turbulent regime.

Needless to say, equations in the class (6.1), (6.2) are much more
general. We mention that for $\a = 2$ we have the generalized
gaussian process $$ u_t \ = \ t^{2 \nu - 1} \ u_{xx} \ . $$

We have the following general result, confirmed (and actually
suggested) by numerical experiments on a number of RL equations
\cite{MVV2}.

\medskip\noindent
{\bf Theorem 1.} {\it The universal asymptotic solution to RL
equations (6.1), (6.2) for $L^1$ initial data is given by
$$ u (x,t) \ \simeq \ {1 \over t^\nu} \exp [- x^\a / t^{2 \nu}] \ . \eqno(6.3) $$}

\medskip\noindent
{\bf Proof.}
In the general case described by (6.1), (6.2), the change of variables
$$ \tau = t^{\a \nu} \ , \ \xi = x^{\a/2} \ , \ w = t^{(2-\a)(\nu/2)} u \eqno(6.4) $$
maps the equation (6.1), (6.2) into the heat equation
$$ w_\tau \ = \ w_{\xi \xi} \ . \eqno(6.5) $$
Using the inverse change of coordinates we have that the universal
asymptotic solution
$$ w (\xi , \tau ) \ \simeq \ \tau^{- 1/2} \ e^{-\xi^2/\tau} \eqno(6.6) $$
of the heat equation is mapped back into the function given in the statement.
This represents therefore the universal asymptotic solution of the equation (6.1), (6.2). \EOP

\medskip\noindent
{\bf Remark 5.} It is appropriate to stress that this result
applies to a quite large class of stochastic processes. This
includes processes with non-independent but finite variance
increments (coloured noise), and  processes with independent
increments but infinite variance (Levy flights); see e.g.
\cite{MVV1,MVV2} again for examples and detailed numerical
studies.

The relevant point is that, in order to apply our approach, the considered stochastic processes should present an asymptotic probability distribution (starting with sufficiently localized initial distribution) of the form
$$ P (x,t) \ \simeq \ \rho t^{-\nu} Q(x/t^\nu) \ \ \ \ (|x| \to \infty \, , \, t \to \infty ) \ , \eqno(6.7) $$
with $\rho$ a normalization constant. E.g., for coloured noise $Q(z) = e^{- z^\a}$. \EOR

\medskip\noindent
{\bf Remark 6.} We also note that one can use this approach for
stochastic processes which are not described in terms of a
diffusion process, but which show in numerical experiments a
limiting distribution of the form (6.7). This is the case e.g. for
the random walk on a comb lattice (in this case $\nu = 1/4$, and
$Q(z) = \exp [- z^\a]$, $\a = 4/3$); or for Levy flights (here $\a
< 3$, $\nu = (\a -1)^{-1} > 1/2$ and $Q(z) = z^{-\a}$), see
\cite{MVV2} for details (the equation studied there in relation to
Levy flights is a phenomenological one, corresponding to the
propagator). In these cases we can use our argument to provide an
effective equation for the limiting behaviour of the probability
distribution; this will provide, by construction, the correct
behaviour in the large $|x|$ and $t$ limit. \EOR

\section{The FKPP equation: asymptotic solution, and symmetries}

We now apply our approach to the
Fisher-Kolmogorov-Petrovskii-Piskunov (FKPP) equation
\cite{Fis,KPP} reads
$$ u_t \ = \ D u_{xx} \, + \, \eps u (1-u) \ ,  $$
with $\eps$ and $D$ real positive constants. Here $D$ represents a diffusion coefficients, while the parameter $\eps$ is the inverse of the timescale for a logistic growth. We are interested in solutions such that $u(x,t) \ge 0$ for all $x$ and $t$.

There are two stationary homogeneous states, i.e. $u = 0$ and
$u=1$; the latter is stable while the former is unstable against
small perturbations.\footnote{The FKPP equation was introduced in
the thirties to model the evolution of the concentration $u(x,t)$
of a dominant gene in a system undergoing a logistic growth (with
time constant $\tau = \eps^{-1}$) and such that between two
reproductive cycles the individuals move around randomly with
diffusivity $D$. The same equation also describes autocatalytic
reactions; in this case $u (x,t)$ represents the concentration of
the products of the reaction itself.}

\subsection{Asymptotic behaviour of solutions}

It is well known -- and it was proved by Kolmogorov, Petrovskii and Piskunov \cite{KPP} -- that if the initial datum is suitably concentrated, e.g. $u (x,0) = 0$ for $|x-x_0| > L$ (i.e. has compact support) as in the cases studied in \cite{MVV2}, or more generally $u(x,0) < A \exp [ - x / L ]$, then asymptotically for $t \to \infty$ and $x \to \infty$ the solution is of the form $ u = f(x,t) \simeq \exp [ - (x - v t) / \la ]$, with $\la = \sqrt{D/\eps}$ and $v = \sqrt{4 \eps D}$. This  represents a {\it front} of width $\la$ travelling with speed $v$; it connects the stable state $u=1$ and the unstable state $u=0$.

In discussing the FKPP equation, it is convenient to pass to rescaled coordinates $ \~t = \eps t$, $\~x = (\sqrt{\eps / D}) x$. From now on we will use these coordinates, and omit the tildas for ease of notation. In these coordinates, the FKPP equation reads
$$ u_t \ = \ u_{xx} + u (1-u) \ . \eqno(7.1) $$
As for the asymptotic solution given above, this reads now
$$u = f_0 (x,t) \ \simeq \ A \ \exp \[ - (x - 2 t) \] \ ; \eqno(7.2) $$
note the front has speed $v=2$ and width $\la=1$.

It should be stressed that the $f(x,t)$ or $f_0 (x,t)$ given above
provide the solution for $x \to \infty$, i.e. in the
region\footnote{Note that for $x\to 0$ and $t\to \infty$ we are in
the region of $u \simeq 1$; writing $u = 1 - \b$ we get in this
region the approximated equation $\b_t = \b_{xx} + 2 \b -1$. With
the ansatz $\b (x,t) = \b (x - 2 t)$, we get the solution $\b (z)
= (1/2) [1 + e^{-z} (c_1 \cos (z) + c_2 \sin (z) )]$, with $c_i$
arbitrary constants.} of small $u$; in this region (7.1) is well
approximated by its linearization around $u=0$, i.e.
$$ u_t = u_{xx} + u \ ; \eqno(7.3) $$
the ansatz $u (x,t) = w(z) := w (x - 2 t)$ takes this into the ODE
$$ w'' + 2 w' + w \ = \ 0  \eqno(7.4) $$
for $w = w(z)$, with solution
$$ w (z) \ = \ c_1 e^{-z} + c_2 z e^{-z} \ . \eqno(7.5) $$

The $f_0$ given above, see (7.2), corresponds to $c_2 = 0$.
This can be characterized in terms of symmetry properties, as
discussed in the next subsection.

\subsection{Symmetry properties of the linearized FKPP equation,
and of its asymptotic solutions}

We will now discuss the symmetry properties of (7.3), (7.4) and of
the solutions (7.5), and characterize the asymptotic solutions in
terms of their symmetry properties; it suffices to consider
translations and scalings.

It is convenient for our discussion to consider linear combinations of the shifts in the $(x,t)$ coordinates, given by $X_\pm = \pa_x \mp (1/2) \pa_t$; note that $z = x - 2 t$ is invariant under $X_-$, and that $X_+ = \pa_z$. We also write $X_0 = u \pa_u$. Needless to
say, $[X_0,X_+] = 0$.

\medskip\noindent
{\bf Lemma 1.} {\it The symmetry algebra of the linearized equation (7.3) is generated by the scaling $X_0$ and by the translations $X_\pm$. The quotient equation (7.4) admits only $X_0$ as scaling symmetry; it also admits the translation symmetry generated by $X_+$, while $X_-$ has been quotiented out by passing to the $z$ variable.}

\medskip\noindent
{\bf Proof.} This follows from direct computations. \EOP

Let us consider the whole set ${\cal W}$ of solutions described by
(7.5); we will denote the solution $c_1 e^{-z} + c_2 z e^{-z}$ as
$\ket{c_1,c_2}$. Note that ${\cal W} = \R^2$, and $(c_1,c_2)$
provide coordinates in ${\cal W}$.

\medskip\noindent
{\bf Lemma 2.} {\it The propagating front solutions are selected among all those in ${\cal W}$ by their symmetry properties; more specifically, they correspond to an invariant subspace of ${\cal W}$ under the action of the group generated by the vector fields $X_0$ and $X_+$.}

\medskip\noindent
{\bf Proof.}
The transformation $e^{\a X_0}$ maps $w$ into $e^{\a} w$ without affecting $z$; the transformation $e^{\b X_+}$ maps $z \to e^\b z$ without affecting $w$. A general element of the group generated by $X_0$ and $X_+$ is written as $g (\a , \b) := \exp [ \a X_0 + \b X_+ ]$ and acts on ${\cal W}$ by
$$ g (\a,\b) \ : \ \ket{c_1,c_2} \ \to \ \ket{ e^{\a + \b} (c_1 + \b c_2 ) , e^{\a + \b} c_2 } \ := \ \ket{\mu (c_1 + \b c_2) , \mu c_2 } \ . \eqno(7.6) $$
In the last equality we have defined $\mu := \exp (\a + \b )$.
Thus the action of $g(\a,\b)$ on ${\cal W}$ is given, in terms of the $(c_1,c_2)$ coordinates, by a matrix
$$ M_{\a,\b} \ := \ \pmatrix{\mu & \b \mu \cr 0 & \mu \cr} \ . \eqno(7.7) $$
The subspace $c_2 = 0$ is invariant under this action: in other
words, the propagating front solutions (7.2) are indeed selected by a
symmetry property as claimed. \EOP

\subsection{Asymptotic symmetry analysis of the FKPP equation}

It is immediate to see that the only scaling or shift symmetries
of the full FKPP equation (7.1) are those, with generators $X_1 =
\pa_x$ and $X_2 = \pa_t$, corresponding to translations in $x$ and
$t$; these reflect the fact that (7.1) is a homogeneous equation.

The situation is different for what concerns asymptotic
symmetries, and in particular scaling ones, as we now discuss.

\medskip\noindent
{\bf Lemma 3.} {\it Let $X$ be a scaling vector field, such that $\lim_{\la \to \infty} \exp (\la X)$ extracts the behaviour for large $|x|$ and $t$. Let $\De_0$ be the FKPP equation, and $\De_\la = e^{\la \De} \De_0$. Then $\lim_{\la \to \infty} \De_\la = \De_*$ is the heat equation $u_t - u_{xx}$.}

\medskip\noindent
{\bf Proof.}
We consider the most general scaling generator, i.e. a vector
field in the form (3.1), $ X = a x \pa_x + b t \pa_t + c u \pa_u$.
We can always set one of the constants $(a,b,c)$ equal to unity
(provided it is nonzero); this amounts to a redefinition of the
scaling group parameter.

Applying the procedure described in previous sections, with of
course $\De_0 := u_t - u_{xx} - u (1 -u) = 0$ the FKPP equation,
we obtain at once that
$$ \De_\la \ = \ \la^{c-b} \ \[ u_t \, - \, \la^{b - 2 a} u_{xx} \, - \, \la^b u \, + \, \la^{b+c} u^2 \] \ . \eqno(7.8) $$
We choose $c=b < 0$ and  $a=b/2$.
As mentioned above, we can set the modulus of one of the constants, say $b$ for definiteness, equal to unity; i.e. $b = -1$. With these choices, we have
$$ \De_\la \ = \ u_t \, - \, u_{xx} \, - \, \la^{-1} u \, + \, \la^{-2} u^2 \ . \eqno(7.9) $$
Note we want to extract the behaviour for large $x$ and $t$; this corresponds to the limit $\la \to \infty$.
The limit $ \De_*  :=  \lim_{\la \to \infty} \De_\la$ is the heat equation $u_t - u_{xx} = 0$, as claimed. \EOP

\section{Richardson-like logistic reaction-diffusion equations}

We now consider reaction-diffusion equations associated to RL
diffusion in the same way as the FKPP equation is associated to
standard diffusion.

These are written as
$$ u_t \ = \ \^L [u] \, + \, h (u) \ ; \eqno(8.1) $$
here $\^L$, see (6.2), is the RL linear operator describing
passive transport of the field $u$, while $h (u)$ describes its
growth. The logistic growth, which we will consider here,
corresponds to choosing (note an overall constant could be
reabsorbed by a rescaling of $u$)
$$ h(u) \ := \ u \, (1 - u) \ . \eqno(8.2) $$

With this choice, and with (6.2), we get the RL logistic (or RLL)
equation
$$ u_t \ = \ {x^{2-\a } \over t^{1 - \nu \a} } \ \[ u_{xx} \, + \, {(2 - \a ) \over 2 x} \, u_x \] \ + \ u \, (1 - u) \ . \eqno(8.3) $$
(The case $\nu = 1/2$ corresponds to standard diffusion, while for $\nu \not= 1/2$ we have indeed anomalous diffusion).
One is usually interested in solutions with initial data $u(x,0)$ which are suitably regular and with compact support.

\subsection{Numerical experiments}

A detailed numerical study of RLL equations (8.3) with such
initial data was conducted in \cite{MVV1,MVV2}. We summarize the
findings of these numerical experiments as follows (see
\cite{MVV1,MVV2} for details):
\begin{itemize}
\item{{\tt (i)}} asymptotically for large $x$ and $t$, the solution is
described by a travelling front with varying speed $c (t)$ and
width $\la (t)$; the form of this front for small $u$ is well
described by
$$ u (x,t) \ \simeq \ A \ \exp \[ - { x - c(t) \cdot t \over \la (t) } \] \ ; \eqno(8.4) $$

\item{{\tt (ii)}} the (asymptotic) scaling properties of $c(t)$ and $\la
(t)$ are described by
$$ c(t) \ \simeq \ c_0 \cdot t^{\de} \ , \ \la (t) \ \simeq \ \la_0 \cdot t^{\de} \eqno(8.5) $$
where $c_0$ and $\la_0$ are dimensional constants;
\item{{\tt (iii)}}
the scaling exponent $\de$ is given by
$$ \de \ := \ \nu \, + \, (1/\a) \, - \, 1 \ . \eqno(8.6) $$
\end{itemize}

Thus we rewrite (8.4) in the form
$$ u(x,t) \ \simeq \ A \ \exp \[ - { x - (c_0 t^\de) t  \over \la_0 t^\de } \] \ .  \eqno(8.7) $$
Note that for $\de = 0$, i.e. for $\nu = 1 - (1/\a)$, the front
travels with constant speed and width, as for the standard FKPP
equation.

\subsection{Asymptotic scaling invariance}

We want now to describe precisely the invariance properties of the observed asymptotic solution (8.7), in particular for what concerns scaling transformations.

\medskip\noindent
{\bf Lemma 4.} {\it The scaling invariance of the function $u(x,t)$ given by (8.7) is described by the generalized scaling group
$$ \cases{
x \to \ (\mu^\de) \ x \ , & \cr
t \to \ \mu \ t \ , & \cr
u \to \ \[ \exp \( {(\mu - 1) \, K \, t } \) \] \ u \ , & \cr}  $$
with $\mu$ the group parameter.}

\medskip\noindent
{\bf Proof.}
We are interested in vector fields of the form
$$ X \ = \ a \, x \, \pa_x \ + \ b \, t \, \pa_t \ + \ \b (x,t) \, u \, \pa_u  \eqno(8.8) $$
which leave (8.7) invariant. It is immediate to note that, in order to leave the functional form of (8.7) invariant -- i.e. to have at most a redifinition of the dimensional constant $A$ -- we must require $a = b \de$ in (8.8); moreover, we can set $b = 1$ with no loss of generality: this amounts to reparametrization of the one-parameter subgroup generated by $X$. Thus we are reduced to considering vector fields of the form
$$ X \ = \ \de \, x \, \pa_x \ + \ t \, \pa_t \ + \ \b (x,t) \, u \, \pa_u \ . \eqno(8.9) $$
The expression of $\b$ is obtained by the requirement that $f(x,t)$ defined in (8.7) is invariant. Applying (1.2) and recalling $\phi (x,t,u) = \b (x,t) \cdot u$ in the present case, this amounts to
$$ \b (x,t) \ f(x,t) \ + \ \de \, x \, (\pa f / \pa x) \ - \ t \, (\pa f / \pa t) \ = \ 0 \ . \eqno(8.10) $$
This is immediately solved: we get $\b (x,t) = - (c_0 / \la_0) t$; for ease of notation, we will write $K := c_0 / \la_0 $, so $\b (x,t) = - K t$. Inserting this in (8.9), we finally obtain that (8.7) is invariant under the scaling-like vector field
$$ X \ = \ \de \, x \, \pa_x \ + \ t \, \pa_t \ - \ K t \, u \, \pa_u \ . \eqno(8.11) $$
This can be easily checked using (1.2): with $u = f(x,t)$ given by (8.7), and writing as usual (8.11) in the form $X = \xi \pa_x + \tau \pa_t + \phi \pa_u$, we get $\de f := \phi = \xi u_x - \tau u_t = 0$.

The one-parameter group generated by $X$ acting on the point $(x_0,t_0,u_0)$ is given by
$$ t (s) = e^s t_0 \ , \ x(s) = e^{\de s} x_0 \ , \ u (s) = \exp \[ {K (t(s) - t_0 ) } \] u_0 \ . \eqno(8.12') $$
This is indeed the generalized scaling group (note that the scaling in $u$ is $t$-dependent) given in the statement. \EOP

\subsection{Partial asymptotic scaling symmetry}

\medskip\noindent
{\bf Theorem 2.} {\it The scaling vector field (8.11) is not a
symmetry of the RLL equation (8.3). It is an asymptotic symmetry
of the same equation.}

\medskip\noindent
{\bf Proof.} Let us denote (8.3) by $\De_0$ and (8.11) by $X$; applying the second prolongation $Y \equiv X^{(2)}$ of (8.7) on $\De_0$, and restricting to the solution manifold $S_0$ of $\De_0$, which amounts to substituting for $w_\s$ according to $\De_0$ itself, we obtain
$$ \begin{array}{rl}
\De_1 := \ \[ Y (\De_0 ) \]_{S_0} \ =& \ [ (1- \a) t^{\a \de} \(x/t\)^{2-\a} ] \, u_{xx} \ + \\
& \ +  \[ (1/2) (\a - 1 ) (\a - 2) t^{\a \de - 1} \( x/t \)^{1 - \a } \] \, u_x \ + \\
& \ - \ u ( 1 + K - u + K u t )  \ . \end{array} \eqno(8.13) $$
This is not zero, i.e. $X$ is not a symmetry of (8.3).

Going further on with our procedure, we apply $Y$ on $\De_1$, and restrict the obtained expression to $S_0 \cap S_1$, with $S_i$ the solution manifold of $\De_i$; we obtain
$$ \begin{array}{rl}
\De_2 \ :=& \[ Y (\De_1 ) \]_{S_0 \cap S_1} \ = \
\[ 2 - 4 K t + K^2 t^2 - \a (1 - K t ) \] \, u^2 \ + \\
& \ - \ \[ (1+K)(\a - 2 ) \] u \ = \ 0 \ . \end{array} \eqno(8.14) $$
Needless to say, this has the trivial solution $u = 0$, which is also solution to $\De_0$ and $\De_1$, and the nontrivial solution
$$ u (t) \ = \ { (2 - \a ) (1 + K ) \over (2 - \a) - (4 - \a ) K t + K^2 t^2 } \ . \eqno(8.15) $$
The latter, as easily checked by explicit computation, is in general {\it not} a solution to $\De_0$ and $\De_1$.
Indeed inserting this into $\De_0$ and $\De_1$ we have respectively
$$ \widetilde{\De_0} \ = \ \[(\a - 2) K (1 + K)\] \ \({K t^2 + (2 K - 4 +  \a ) t + (2 \a  - 6 ) \over \( K^2 t^2 + (\a - 4 K) t + (2 - \a ) \)^2 } \) \ , \eqno(8.16') $$
$$ \widetilde{\De_1} \ = \ \[ (\a - 2) K (1+K)^2 \] \ \( { K t^2 - 2 t \over \( K^2 t^2 + (\a - 4 K ) t + (2 - \a ) \)^2 } \) \ . \eqno(8.16'') $$
For $K \not= 0,-1$, both of these expressions are not zero (unless
$\a = 2$). However, both of these go to zero like $(1/t^2)$, for
all $\a$ and $K$, in the limit $t \to \infty$. This concludes the
proof.\footnote{We could of course also have worked with the
symmetry adapted $(\s , y ; w)$ variables introduced in the next
subsection, i.e. with the expression (8.18) for $X$ and (8.20) for
$\De_0$, reaching the same result.} \EOP

\subsection{Solution-preserving map associated to scaling}

Having determined that $X$ is an asymptotic (partial) symmetry for our equation $\De_0$, we will now apply our general procedure described in section 5 and look for asymptotically $X$-symmetric solutions to $\De_0$.

The first step consists in passing to symmetry adapted coordinates and expressing the equation in these.

The change to symmetry-adapted coordinates and its inverse are in this case given by
$$ \begin{array}{lll}
\s = t \ , \ & y = x / t^\de \ , \ & w = u e^{K t } \ ; \\
t = \s \ , \ & x = \s^\de y \ , \ & u = w e^{- K \s } \ . \end{array} \eqno(8.17') $$

In the "new" coordinates, the vector field (8.11) reads simply
$$ X \ = \ \s \, \pa_\s \ , \eqno(8.18) $$
and the (obviously $X$-invariant) asymptotic solution (8.7) is
$$ w \ = \ A \ \exp \[ { y \over \la_0 } \] \ . \eqno(8.19) $$

As explained above, once we identify $w$ as the new dependent variable, i.e. $w = w(\s , y)$, the change of variables (8.17') is prolonged to a relation between partial derivatives. With standard computations we obtain
$$ \begin{array}{rl}
u_t \ =& \ \( w_\s - \de (y / \s ) w_y - K w \) \ e^{- K \s } \ , \\
u_x \ =& \ (1 / \s^\de ) \, w_y \ e^{- K \s } \ , \\
u_{xx} \ =& \ (1 / \s^\de )^2 \, w_{yy} \ e^{- K \s } \ . \end{array} \eqno(8.17'') $$

Inserting (8.17) in the expression (8.3) for $\De_0$, we obtain the expression for the latter in the new coordinates; this results to be
$$ w_\s \, = \, \[ {y^{2 - \a} \over \s^{\chi} } \] \ w_{yy}  +  \[ \( {2 - \a \over 2} \)  \( {y^{1-\a} \over \s^{\chi} } \)  +  \a \({y \over \s} \) \] \ w_y  + (K+1) w - e^{- K \s} w^2 \ , \eqno(8.20) $$
where we have written $\chi = \a (\de - \nu + 1/ \a )$ for ease of writing.

The expression (8.20) holds for the general map (8.17); however we are specially interested in the choice $\de = (\nu -1 + 1/\a )$, see (8.6). With this, we have $\chi = (2 - \a)$, and finally (8.3) reads
$$ w_\s \ = \ \( { y \over \s} \)^{2 - \a} \, w_{yy} \ + \ \[ \( {2 - \a \over 2 \s } \) \({y \over \s }\)^{1-\a} + \a  { y \over \s } \] \, w_y \ + \ (K+1) \, w \ - \ e^{- K \s } \, w^2 \ . \eqno(8.21) $$

Note that in the limit $\s \to \infty$, the last term disappears (faster than any power in $\s$), and (8.21) reduce to a linear equation.

\subsection{Scaling-invariant (asymptotic) solutions}

\medskip\noindent
{\bf Theorem 3.} {\it The equation (8.21), has no nontrivial $X$-invariant solutions. It admits nontrivial asymptotically $X$-invariant solutions.}

\medskip\noindent
{\bf Proof.}
The $X$-invariant solutions to (8.21) are obtained by requiring that $w_\s = 0$; with this the equation reduces to
$$ \( { y \over \s} \)^{2 - \a} \, w_{yy} \ + \ \[ \( {2 - \a \over 2 \s } \) \({y \over \s }\)^{1-\a} + \a  { y \over \s } \] \, w_y \ + \ (K+1) \, w \ - \ e^{- K \s } \, w^2 \ = \ 0 \ . \eqno(8.22) $$
Note that $\s$ appears parametrically here, and (8.22) splits into
the equations corresponding to the vanishing of coefficients of
different powers of $\s$ (this is a general feature of partial or
"weak" symmetries, see \cite{CiK,OlV}). The only common solution
to these is $w = 0$, which ends the proof of the first part or the
statement.

We go back to considering (8.21); in order to study its asymptotic behaviour for $\s \to \infty$, we disregard the term which is exponentially small for large $\s$. The resulting linear equation for $w = w(y)$, i.e.
$$ \( { y \over \s} \)^{2 - \a} \, w_{yy} \ + \ \[ \( {2 - \a \over 2 \s } \) \({y \over \s }\)^{1-\a} + \a  { y \over \s } \] \, w_y \ + \ (K+1) \, w \  = \ 0 \ , \eqno(8.23) $$
yields as solution
$$ w(y) \ = \ C_1 \, \K [ 0 , 0 ]  \, + \, \( \sqrt{ 2 (K+1) \s^2 (y/\s)^\a } \) \, C_2 \, \K [1/2 , 1 ] \eqno(8.24) $$
where $C_1 , C_2$ are arbitrary constants, and
$$ \K [ x,y ] \ := \ F_{11} \[ (K+1) \s / \a^2 + x ; 1/2 + y ; - \s (y/\s)^\a \] \eqno(8.25) $$ with $F_{11} \equiv {}_1F_1$ the Kummer confluent hypergeometric function
$$ F_{11} [a;b;z] := \, {}_1F_1 (a;b;z) \ = \ { \Ga (b) \over \Ga (b-a) \, \Ga (a) } \ \int_0^1 \, e^{z t} \, t^{a-1} \, (1-t)^{b-a-1} \, {\rm d} t \ . \eqno(8.26) $$

Needless to say, the asymptotic solution (8.24) could be expressed in terms of the original variables using (8.17); this yields an explicit but involved and not specially illuminating expression. \EOP

\section{Other asymptotic partial symmetries of RLL equations}

In the previous section we have analyzed the RLL equation (8.3) on
the basis of the scaling symmetry (8.11) of its asymptotic
solution (8.7). However, this is not the only symmetry of the
observed asymptotic solution (8.7). In this section we apply our
approach on the basis of different symmetries.

\subsection{General symmetries of the asymptotic solution}

{\bf Lemma 5.} {\it The vector field $X = \xi \pa_x + \tau \pa_t +
\phi \pa_u $ is a symmetry of the asymptotic solution (8.7) to the
RLL equation (8.3) if and only if it belongs to the two
dimensional module (over smooth real functions ${\cal C}^\infty
(\R^3 , \R)$ of $x,t,u$) generated by
$$ \begin{array}{rl}
X_1 \ =& \ \pa_x + \( {1 \over \la_0 t^\de } \) \pa_u \ ; \\
X_2 \ =& \ \( \la_0 t^{1 + \de } \) \pa_t - \( x \de + c_0 t^{1 + \de} \) \pa_u \ . \end{array} \eqno(9.1) $$}

\medskip\noindent
{\bf Proof.} This follows easily by using (1.2) and the explicit expression (8.7) of the asymptotic solution $u = f_* (x,t)$. Indeed, applying (1.2) we get
$$ {(\~f - f ) \over \eps } \ = \ \phi \, - \, A \, \exp \[ (c_0 t - x t^{- \de})/\la_0 \] \, { (x \de + c_0 t^{1 + \de} ) \tau - t \xi \ over \la_0 t^{1 + \de } } \ , $$
and the result follows immediately \EOP

In the following, we will consider in particular\footnote{Note that $u$ is not acted upon by this, i.e. we have a vector field which is horizontal for the fibration $(M,\pi,B)$.}
$$ X_0 \ := \ \( x \de + c_0 t^{1+\de} \) \pa_x + t \pa_t \ ,  \eqno(9.2) $$
as well as $X_1$ and $X_2$ themselves.

We will write second-prolonged vector fields in the form
$$ Y \equiv X^{(2)} \ = \ X \ + \ \Psi_x {\pa \over \pa u_x} + \Psi_t {\pa \over \pa u_t} + \Psi_{xx} {\pa \over \pa u_{xx}} + \Psi_{xt} {\pa \over \pa u_{xt}} + \Psi_{tt} {\pa \over \pa u_{tt}} \ ;  \eqno(9.3) $$
we will actually need only the coefficients $\Psi_x , \Psi_t , \Psi_{xx} $.

We will, as in the previous discussion, denote (8.3) as $\De_0$.
Let us start by considering $X_1$.

\medskip\noindent
{\bf Theorem 4.} {\it The vector fields $X_0$, $X_1$ and $X_2$ are  partial symmetries of $\De_0$.}

\medskip\noindent
{\bf Proof.} We will denote by $Y_i$ the second prolongations of $X_i$. Let us start by considering $X_1$.
In this case the coefficients of the second-prolonged vector field $Y_1$ are:
$$ \Psi_x = 0 \ , \ \Psi_t = - (\de / \la_0) t^{-(1+\de)} ) \ , \ \Psi_{xx} = 0 \ , \ \Psi_{xy} = 0 , \Psi_{tt} = (\de / \la_0 ) (1 + \de) t^{-(2 + \de )} \ . \eqno(9.4) $$
We define then
$$ \begin{array}{l}
\De_1 \ = \ \[ Y_1 (\De_0) \]_{S_0} \ ; \\
\De_2 \ = \ \[ Y_1 (\De_1) \]_{S_0 \cap S_1} \ . \end{array} \eqno(9.5) $$
By explicit computation, it results that
$$ \[ Y_1 (\De_2) \]_{S_0 \cap S_1 \cap S_2} \ = \ 0 \ ; $$
this shows that $X_1$ is a partial symmetry for $\De_0$.

Let us now consider $X_2$. The relevant coefficients of $Y_2$ are
$$ \Psi_x = - {\de \over \la_0 t^{1 + \de }} \ , \ \Psi_t = {(1 + \de ) \de x \over \la_0 t^{2 + \de} } \ , \ \Psi_{xx} = 0 \ . \eqno(9.6) $$
We get more involved expressions for $\De_1$ and $\De_2$ obtained as above (we mention that in this case, $\De_2$ identifies a single function $u = u(x,t)$), but again
$$ \[ Y_2 (\De_2) \]_{S_0 \cap S_1 \cap S_2} \ = \ 0 \ ; $$
this shows that $X_2$ is also a partial symmetry for $\De_0$.

Finally, we consider the vector field $X_0$, in which we are specially interested. The relevant coefficients of $Y_0$ are
$$ \Psi_x = - \de u_x \ , \ \Psi_t = - (1 + \de ) c_0 t^\de u_x - u_t \ , \ \Psi_{xx} = - 2 \de u_{xx} \ . \eqno(9.7) $$
Defining $\De_1 = Y_0 (\De_0)$ and $\De_2 = Y_0 (\De_1)$, we obtain again that
$$ \[ Y_0 (\De_2) \]_{S_0 \cap S_1 \cap S_2} \ = \ 0 \ , $$
hence $X_0$ is also a partial symmetry for $\De_0$. \EOP

\subsection{Invariant solutions}

It turns out that reduction, and invariant solutions, under the vector field $X_0$ are of special interest. This is due to the following theorem, which provides an analytic explanation of the numerically observed behaviour.

\medskip\noindent
{\bf Theorem 5.} {\it The RLL equation (8.3) admits an
asymptotically $X_0$-invariant solution, described by (8.7).}

\medskip\noindent
{\bf Proof.} In this case the symmetry adapted coordinates are
$$ \s = t \ , \ y = (x / t^\de ) - c_0 t \ , \ w = u \ . \eqno(9.8) $$
The inverse change of coordinates is therefore given by
$$ t = \s \ , \ x = (y + c_0 \s ) \s^\de \ , \ u = w \eqno(9.9) $$
and the relevant $u$ derivatives are expressed in the new coordinates as
$$ u_t = w_\s - \( { \de y + c_0 (1 + \de ) \s \over \s } \) w_y \ , \ u_x = {1 \over t^\de} w_y \ , \ u_{xx} = {1 \over t^{2 \de}} w_{yy} \ . \eqno(9.10) $$
The equation (8.3) is therefore written, in these coordinates, as
$$ \begin{array}{rl}
w_\s  =& \ \({ c_0 (1 + \de) \s + \de y \over \s} \) w_y \, + \, \( { (c_0 \s + y)^{1-\a}  \over 2 \s^{\a (\de - \nu ) + 1} } \) \, \[ (2 - \a ) w_y + 2 (c_0 \s + y ) w_{yy} \] \\
 & \ + \ (1-w) w \ . \end{array} \eqno(9.11) $$
In these coordinates $X_0$ reads simply
$$ X_0 \ = \ \s \pa_\s \ , \eqno(9.12) $$
and its second prolongation is
$$ Y_0 \ = \ \s {\pa \over \pa \s} - w_\s {\pa \over \pa w_\s} - w_{\s y} {\pa \over \pa w_{\s y} } - 2 w_{\s \s} {\pa \over \pa w_{\s \s} } \ . \eqno(9.13) $$

In order to discuss (9.11), it is convenient to rewrite it as
$$ w_\s \ = \ A \, w_{yy} \, + \, B \, w_y \, + \, f(w) \ , \eqno(9.14) $$
with
$$ A \ = \ \( {y+c_0 \s \over \s}\)^{2-\a } \ \ ; \ \
B \ = \ \( {c_0 + \de {y + c_0 \s \over \s} + \epsilon \( {y + c_0 \s \over \s } \)^{2-\a} {1 \over y+c_0 \s}} \) \ . \eqno(9.15) $$

For $\s \to \infty$, (9.14) reads
$$ w_\s \ = \ c_0^{2-\a} \, w_{yy} \ + \ c_0 (1+\de) \, w_y \ + \ f(w) \ . \eqno(9.16) $$
The $X_0$-invariant solutions satisfy $w_\s=0$, i.e. $w = w(y)$, and are thus obtained as solution to
$$ c_0^{2-\a} w_{yy} \ + \ c_0 (1+\de) w_y \ + \ w \ =\ 0 \ , \eqno(9.17) $$
where we have used $w << 1$ in the region we are investigating, i.e. for $\s \to \infty$, so that $f(w) \simeq w$.

Solutions to (9.17) are of the form
$$ w (y) \ = \ c_1 e^{-\om_+ y} \ + \ c_2 e^{-\om_- y} \ , \eqno(9.18) $$
where
$\om_\pm = { (1 +  \de) \over 2 c_0^{1-\a} } \[ {1 \pm \sqrt{1-{4 \over c_0^\a (1+\de)^2}} } \]$.
If we require the solutions to be non oscillating, this implies a lower bound on the parameter $c_0$, i.e. $c_0 \ge (2/(1+\de))^{2 /\a}$.

The solution $e^{-\om_+ z}$ is unstable against small perturbations, while $e^{-\omega_- z}$ is stable \cite{KPP}. As proved by Kolmogorov \cite{KPP}, the asymptotic solution is the stable one with the lowest speed giving nonoscillating behaviour, i.e. $ c_0 = [ 2 / (1 + \de) ]^{2 /\a}$. This means
$ w(y) \simeq e^{- \om_0 y }$ with $\om_0 = [2 / (1+\de)]^{1-2/\a}$.
Going back to the original variables, we get
$$ u(x,t) \ \simeq \ A \ \exp{\[ {-{x-v(t) t \over \la (t)}}\]} \ = \
A \ \exp \[ - \om_0 {{x-c_0 t^{1+\de} \over t^\de}} \] \ . \eqno(9.19) $$
This is precisely the numerically observed asymptotic behaviour, described by (8.7); see \cite{MVV2}. \EOP

\section{Asymptotic symmetry in an optical lattice}

In this section we apply our approach to a different kind of
anomalous diffusion equations; that is, we focus on the equation
describing anomalous transport in an optical lattice \cite{Lut}.
The equation for the marginal Wigner distribution $w(p,t)$ of the
momentum $p$ at time $t$ reads
$$ w_t \ = \ - {\pa \over \pa p } \ \[ h(p) \, w \ - \ g(p) \, w_p \] \eqno(10.1) $$
where the functions $h(p)$ and $g(p)$ are given by
$$ h (p) \ := \ {\a p \over 1 + (p / p_c)^2 } \ \ , \ \ g(p) \ := \ \ga_0 \ + \ {\ga_1  \over 1 + (p / p_c)^2 } \ ; \eqno(10.2) $$
here $\a , \ga_0 , \ga_1 $ are certain constants, and $p_c$
represents the capture momentum.

With the shorthand notation $ \b (p ) := 1/[1 + (p/p_c)^2 ] $, the
equation (10.1), (10.2) reads
$$ \begin{array}{rl}
w_t \ =& \ \( \ga_0 + \b (p) \ga_1 \) \, w_{pp} \ + \
\( \a \b (p) \, - \, 2 \ga_1 [\b (p) / p_c ]^2 \) \, p \, w_p \ + \\
 & \ + \ \a \, \( \b (p) \, - \, 2 [ (p/p_c) \b (p) ]^2 \) \, w \ . \end{array} \eqno(10.3) $$

Attention to eq. (10.3) was recently called by Lutz \cite{Lut} (to
which the reader is also referred for derivation and a discussion
of this equation), who remarked that -- quite surprisingly -- the
equilibrium distribution is related to Tsallis statistics
\cite{Tsa}. Indeed, for $q<3$ the stationary solution of (10.3) is
the Tsallis distribution
$$ w_0 (p) \ = \ {1 \over Z} \ \[ 1 \, - \, (\b  / \mu ) \, p^2 \]^\mu \eqno(10.4) $$
where
$$ \b  =  {\a \over 2 (\ga_0 + \ga_1 )} \ , \ \mu = {1 \over 1 - q} \ , \ q = 1 + {2 \ga_0 \over \a p_c^2} := 1 + \delta \ , $$
and $Z$ is a normalization factor, which for $q<3$ can be chosen
so that $\int_{- \infty}^{\infty} w_0 (p) \d p = 1 $. For $5/3 < q
< 3$ the second moment $\int p^2 w_0 (p) \d p$ of the Tsallis
distribution is infinite, and we have anomalous diffusion.

Here we will consider generalized scaling
transformations\footnote{The peculiar properties of (10.3) and its
stationary solution, in particular concerning the scale-free
nature of $w_0 (p)$, has been studied by Abe \cite{Abe} in
connection with dilation symmetries and canonical formalism. Note
however that he introduces a canonical structure, which we do not
need here, and that his symmetry generators are nonlocal, being
expressed as integrals over $p$. Thus, our approach is
substantially different from his one.}; that is, transformations
acting as a standard scaling in the independent $p$ and $t$
variables, and as a $p$ and $t$ dependent scaling in the dependent
variable $w(p,t)$:
$$ X \ = \ c_1 p \pa_p + c_2 t  \pa_t + \phi (p,t,w) \pa_w \ . \eqno(10.5) $$
We will of course consider asymptotic invariance.

Applying (1.2) to (10.4) and to generalized scalings we get with
easy computations that (up to a common factor) the transformation
generated by (10.5) with $c_1 = - 1 $ leaves (10.4) invariant if
and only if
$$ \phi \ = \ (2 \, \b \, Z^{q-1} ) \, p^2 \, w^q \ ; \eqno(10.6) $$
obviously $c_2$ remains unrestricted at this stage. We will, for
ease of notation, write
$$ \nu \ := \ 2 \, \b \, Z^{q-1} \ . $$

Let us now discuss what is special in the vector fields
$$ X \ = \ - p \pa_p \ + \ \s t \pa_t \ + \ [\nu \, p^2 \, w^q ] \pa_w \eqno(10.7) $$
for what concerns their action on the equation (10.3).

We are interested in
$$ Y \ = \ X \ + \ \Psi_t {\pa \over \pa w_t} \ + \ \Psi_p {\pa \over \pa w_p} \ + \ \Psi_{pp} {\pa \over \pa w_{pp}} $$
(other prolongation coefficients are irrelevant for application to
our  equation); the coefficients corresponding to (10.7) are
$$ \begin{array}{l}
\Psi_t \ = \ - \s w_t + \nu q p^2 w^\de \ ; \ \ \
\Psi_p \ = \ w_p + 2 \nu p w^{(1+\de)} + \nu (1+\de) p^2 w^\de \ ; \\
\Psi_{pp} \ = \ 2 w_{pp} + 2 \nu w^{(1+\de)} + \nu w^{(\de-1)} \[
\de (1 + \de) p^2 w_p^2 + (1 + \de) p w (4 w_p + p w_{pp} ) \] \ .
\end{array} $$

Applying $Y$ on (10.3), and substituting for $w_t$ according to
(10.3) itself, we obtain an expression which we write as
$$ \De \ = \ \Ga_0 (p,t,w) \ + \ \Ga_1 (p,t,w) \, w_p \ + \ \Ga_2 (p,t,w) \, w_{pp} \ . \eqno(10.8) $$

The explicit expression of $\Ga_2$ is
$$ \Ga_2 \ = \
-{\frac{\ga_0\,( 2 + \s ) \,
       {{( {{p_c}^2} + {p^2} ) }^2} +
      \ga_1\,{{p_c}^2}\,
       ( {{p_c}^2}\,( 2 + \s )  +
         ( 4 + \s ) \,{p^2} ) }{{{( {{p_c}^2} + {p^2} ) }^2}}} \ . $$
The limit of $\Ga_2$ for large $|p|$ is nonzero unless we choose
$$ \s \ = \ - 2 \ . \eqno(10.9) $$
We assume this from now on. With (10.9), $X$ reads
$$ X \ = \ - p \, \pa_p \ - \ 2 \, t \, \pa_t \ + \ \nu \, p^2 \, w^q \, \pa_q \eqno(10.10) $$
and $\Ga_2$ reduces to
$$ \Ga_2 \ = \ {\frac{-2\,\ga_1\,{{p_c}^2}\,{p^2}}
   {{{( {{p_c}^2} + {p^2} ) }^2}}} \ ; $$
therefore, $\Ga_2 \to 0$ for $|p| \to \infty$.

As for $\Ga_1$, with (10.9) it reads
$$ \begin{array}{l} \Ga_1 \ = \
\[ 2 p ( ( p_c^2 + p^2 )
( \a p_c^4 - 2 ( 1 + \delta )  \ga_0 \nu w^\delta
( p_c^2 + w^2 )^2 ) \right.  \\
\left. - 2 \ga_1 p_c^2 ( ( 1 + \delta )  \nu p_c^4 w^\delta + p^2
( -1 + \nu w^\delta p^2 +
\delta \nu w^\delta p^2 ) \right.  \\
\left. + p_c^2 ( 1 + 2 \nu w^\delta p^2 + 2 \delta \nu w^\delta
p^2 )   \] \ \times \[ ( p_c^2 + p^2 )^3
\]^{-1} \ . \end{array} $$ It can be checked that in the limit
$|p| \to \infty$ and $w \to 0$, we get $\Ga_1 \to 0$.

Finally, with (10.9) $\Ga_0$ reads
$$ \begin{array}{rl}
\Ga_0 \ =& \
\[ w ( -2 \nu w^\de
( p_c^2 + p^2 ) ( \ga_1 p_c^2 ( p_c^2 - p^2 )  + \ga_0 ( p_c^2 +
p^2 )^2
)  + \right. \\
 & \left. + \a p_c^2
( -( ( 2 + \delta )  \nu w^\de p^6 )  - 2 p_c^2 p^2
( 3 + 2 \nu w^\de p^2 )  + \right. \\
 & \left. + p_c^4
( 2 - 2 \nu w^\de p^2 + \delta \nu w^\de p^2 ) ) )
\] \ \times \
\[ p_c^2 + p^2 \]^{-3} \ . \end{array} $$
Once again, in the limit $|p| \to \infty$ and $w \to 0$ we get
$\Ga_0 \to 0$.

This explicit computation suggests that
$$ X \ = \ - p \pa_p \ - 2 t \pa_t \ + \ [(2 \, \b \, Z^{q-1} ) \, p^2 \, w^q ] \pa_w \eqno(10.11) $$
is an asymptotic symmetry for the equation (10.3), provided this
is set in the function space identified by $\lim_{|p|\to \infty}
w(p,t) = 0$.\footnote{More precisely, we should also check that
the variation $\de w = p^2 w^q (p,t)$ is normalizable. This is a
more restrictive requirement, but still verified for functions
near to the Tsallis distribution (10.4) for all $q$ in the
physical range $1 < q < 3$. However, this limitation will be
removed below.}

This limitation is removed by a more careful analysis, using the
systematic procedure described above; we are now going to describe
this approach.

Symmetry adapted variables for $X$ are given by
$$ y = p^2/t \ , \ v = w^{- \de } - (\nu \de / 2) p^2 \ , \ \s = t \ ; \eqno(10.12) $$
the inverse change of coordinates is
$$ t = \s \ , \ p = \sqrt{y \s} \ , \ w = (v + (\nu \de /2 ) y \s )^{- 1 / \de } \ . \eqno(10.13) $$
In the new coordinates the vector field (10.11) is simply $ X = -
2 \s \pa_\s$. Its second prolongation is $ Y = X + 2 v_\s (\pa /
\pa v_\s )$.

With standard computations (see sect.1), the partial derivatives
appearing in (10.3) are written in the new coordinates as
$$ \begin{array}{rl}
w_t \ =& \ w_\s - (y/\s) w_y \\
 =& \ - (1/(\de \s)) \[ (\de \nu \s y /2 ) + v \]^{-q/\de} \ \[ \s v_\s - y v_y \] \ , \\
w_p \ =& \ 2 \sqrt{y/\s} w_y \\
 =& \  - (2 \sqrt{y/\s}/\de ) \[ (\de \nu \s y / 2) + v \]^{- q / \de } \ \[ (\de \nu \s / 2) + v_y \] \ , \\
w_{pp} \ =& \ (2/\s) w_y + (4 y/\s) w_{yy} \\
 =& \ 2 \left[ -2 \de v ( \de \nu \s +
2 v_y + 4 y v_{yy} )  + \right. \\
& \left. \ + y ( 2 \de ( 4 + 3 \de ) \nu \s v_y + 8 ( 1 + \de)
v_y^2 + \de^2 \nu \s ( ( 2 + \de ) \nu \s - 4 y v_{yy} )
)  \right] \times \\
& \times  \[ \de^2 \s (\de \nu \s y + 2 v)^2 \]^{-1} \ .
\end{array} \eqno(10.14) $$

Using these we can rewrite the equation (10.3) in $X$-adapted
coordinates; this is a rather involved expression, and we write it
as
$$ \De_0 \ = \ \( \chi_1 v_\s + \chi_2 v_{yy} + \chi_3 v_y^2 + \chi_4 v_y + \theta \) \, \( v +(1/2) \de \nu \s y \)^{- 1/\de} \ ,  $$
where $\chi_i $ and $\theta$ are function of $\s,y,v$ alone, given
by
$$ \begin{array}{rl}
\chi_1 \ =& \ -  2 \, \[ \de (2 v + \de \nu \s y ) \]^{-1} \ , \\
\chi_2 \ =& \  8 y \( (\ga_0 + \ga_1) p_c^2 + \ga_0 \s y \) \, \[ \de \s (p_c^2 + \s y ) ( 2 v + \de \nu \s y ) \]^{-1} \ , \\
\chi_3 \ =& \ 16 y (1 + \de ) \( (\ga_0 + \ga_1 ) p_c^2 + \ga_0 \s y \) \, \[ \de^2 s (p_c^2 + \s y) (2 v + \de \nu \s y )^2 \]^{-1}  \ , \\
\chi_4 \ =& \ -2 \( -  y ( p_c^2 + \s y ) ( p_c^2 - 2 \a p_c^2 \s
+ \s y )
( 2 v + \de \nu \s y )  + \right. \\
& \ + \ \left. 2 \ga_0 ( p_c^2 + \s y )^2
( -2 v + ( 4 + 3 \de ) \nu \s y )  + \right. \\
 & \ + \ \left.
2 \ga_1 p_c^2 ( -2 p_c^2 v + 4 \nu p_c^2 \s
 y + 3 \de \nu p_c^2 \s y + 2 \s v y +
4 \nu \s^2 y^2 + 5 \de \nu \s^2 y^2 )
\) \ \times \\
 & \times \ \[ \de \s ( p_c^2 + \s y )^2 ( 2 v + \de \nu \s y )^2 \]^{-1}  \\
\theta \ =& \ - \[ 2 \ga_0 \nu ( p_c^2  + \s y )^2
( -2 v + ( 2 + \de )  \nu \s y )  + \right. \\
& \ + \ \left. p_c^2  \( \a ( 2 v + \de \nu \s y ) \( -2 p_c^2  v
+ 2 \nu p_c^2  \s y - \de \nu p_c^2  \s y + 2 \s v y +
2 \nu \s^2 y^2 + \de \nu \s^2 y^2 \)  + \right. \right. \\
& \left. \left. \ + \ 2 \ga_1 \nu ( -2 p_c^2  v + 2 \nu p_c^2  \s
y + \de \nu p_c^2  \s y + 2 \s v y + 2 \nu \s^2 y^2 + 3 \de \nu
\s^2 y^2 )
\) \] \ \times \\
 & \ \times \  \[ ( p_c^2  + \s y )^2
( 2 v + \de \nu \s y )^2 \]^{-1} \ . \end{array} $$

Applying $Y$ on (10.3), we obtain a quite complex expression,
which we write as
$$ \De_1 = Y (\De_0) \ = \ \[ \Xi_1 v_{yy} + \Xi_2 v_{y}^2 + \Xi_3 v_y + \Theta \] \ \( v + (1/2) \de \nu \s y \)^{-1/\de} \eqno(10.15) $$
where $\Xi_i$ and $\Theta$ do not depend on $v_y , v_{yy}$. The
explicit expressions of these are not simple, so we will provide
only their series expansion around $s = \infty$ in terms of $\eps
= 1/s$. These are:
$$ \begin{array}{rl}
\Xi_1 \ =& \ 16 \, \[ \( ( 1 + 2 \de ) g_0 \) \, / \, \(\nu \de^3 \)\] \, \eps^2 \ + \ O (\eps^3 ) \ ; \\
\Xi_2 \ =& \ - \, 32 \, \[ (1 + \de ) (1 + 3 \de ) g_0 \, / \,  (\nu^2 \de^5  y ) \]  \, \eps^3 \ + \ O (\eps^4 ) \ ; \\
\Xi_3 \ =& \ 4 \, \[ \( (1 + 2 \de) [\de (y - 2 \a p_c^2) - 2 (4 + 3 \de )] g_0 \) \, / \,  (\nu \de^4  y ) \] \, \eps^2 \ + \ O (\eps^3 ) \ ; \\
\Theta \ =& \ 2 \, \[ \( (2 + 3 \de + \de^2) (2 g_0 + \a \de p_c^2
) \nu \, + \, 2 (1 + 2 \de ) \, v_\s \) \, / \,  \( \nu \de^3  y
\) \] \, \eps \ + \ O (\eps^2 ) \ . \end{array} \eqno(10.16) $$
Needless to say, these all vanish for $\eps \to 0$, i.e. for $s
\to \infty$. This computations shows that:

\medskip\noindent
{\bf Theorem 6.} {\it The one-parameter group of transformations
generated by
$$ X \ = \ - p \pa_p \ - 2 t \pa_t \ + \ [(2 \, \b \, Z^{q-1} ) \, p^2 \, w^q ] \pa_w $$
is not an exact symmetry of the equation (10.3), but is an
asymptotic symmetry for this equation. No vector field of the form
(10.7) with $\s \not= -2$ is an asymptotic symmetry of (10.3).}
\medskip

In other words, the distribution (10.4) is an invariant function
for a transformation $X$, given by (10.10), which is an asymptotic
symmetry of (10.3), and hence the scale-free nature of the
asymptotic solution (10.4) is a consequence of the asymptotic
symmetry properties of the equation.

\section{Discussion and final remarks}

Let us briefly summarize the main findings of previous sections.

In sections 1-2 we recalled some basic facts about geometry of
PDEs and the concepts of full, conditional and partial symmetries
of these; in sections 3-4 we developed our approach, based on
geometrical considerations, to asymptotic versions of these. In
sect.5 we described a method to test the asymptotic behaviour of
nonlinear PDEs based on asymptotic symmetry properties. The rest
of the paper is devoted to application of the general method
described and developed in sects.1-5 to some specific classes of
PDEs.

First of all we considered, in sect.6, a model class of anomalous
diffusion equations, which we called ``Richardson-like'' (RL), see
(6.2), which had been previously studied numerically -- in
particular for what concerns their asymptotic behaviour -- in
detail \cite{MVV2}. With theorem 1, we proved that the asymptotic
properties can be recovered by the standard ones for the heat
equation, via the change of variables (6.3), and are described by
(6.6).

In sect.7 we applied our approach to the standard FKPP equation,
and described in detail its asymptotic symmetry properties, see
lemmas 1-3; we have also shown that our approach recovers the well
known asymptotic properties of FKPP solutions.

In the following two sections we have then applied our method to
anomalous reaction-diffusion equations associated to RL type, also
called ``Richardson-like logistic'' (RLL) equations, which were
also studied numerically in \cite{MVV2}. In sect.8 we recalled the
features of asymptotic solutions as observed in numerical
experiments, and identified the Lie generator $X$, given by
(8.11), of the observed asymptotic generalized scaling invariance
(lemma 4); in theorem 2 we showed that this is {\it not} a
symmetry of the RLL equation, but it is an asymptotic symmetry for
it. We have then considered the solution-preserving maps
associated to this asymptotic scaling symmetry, focusing on the
physical value of the parameter $\de$; in theorem 3 we have shown
that in this case our equation has no solution invariant under $X$
(which therefore is not a conditional symmetry of the equation),
but admits solutions which are {\it asymptotically} invariant
under it: $X$ is an asymptotic conditional symmetry of the model
anomalous RD equation we consider.

In sect.9 we passed to consider in more detail the numerically
observed  asymptotic solution (8.7). With lemma 5 we identified
the full symmetry algebra $\G$ of it; vector fields in this
algebra are asymptotic conditional symmetries of our model RD
equation. In theorem 4 we focused on certain vector fields in
$\G$, and shown they are partial symmetries for the model RD
equation. Among these vector field is the scaling vector field
$X_0$ given by (9.2); it does not depend on, nor acts upon, the
dependent variable $u$. With theorem 5, we proved that the RLL
equation does admit an asymptotically $X_0$-invariant solution,
which is precisely the observed one (8.7).

Finally, in sect.10 we consider a different example of anomalous
diffusion, occurring in optical lattices. This attracted some
attention due to the appearance of a Tsallis distribution as
stationary states in a certain range of parameters. The surprising
fact is that such a distribution decays polynomially, and
therefore is scale-free. Our method can deal successfully with
this equation, and indeed theorem 6 identifies the relevant
asymptotic symmetries. Moreover, the observed (scale-free)
properties of the solution in the physically relevant range can be
explained on the basis of the asymptotic symmetry properties of
the underlying equation.

\subsection*{Final remarks}

Let us present some final remark.
\medskip

1.) One could wonder what is the advantage of the discussion
conducted in sect.9, and in particular of theorem 5, over the one
of sect.8 and theorem 3. The answer is in the properties of $X_0$:
it identifies characteristics (in the sense of the elementary
theory of linear or quasilinear PDEs) in the space of independent
variables $(x,t)$, and reconducts the observed asymptotic
properties of solutions to the RLL equations to their asymptotic
constancy along these characteristics. This feature should not be
surprising: indeed we had already observed in sect.8, cf. eqs.
(8.22) through (8.23), that the asymptotic regime is described by
a quasilinear equation. Our result could then be also described in
terms of "asymptotic characteristics".
\medskip

2.) Our general method to investigate and predict asymptotic
symmetry -- in particular, scaling and front-like -- properties of
solutions to scalar PDEs, represents an evolution of the classical
method to determine partially invariant solutions for symmetric
PDEs \cite{Olv1,Win}, and a blend of it with the method of
conditional and partial symmetries \cite{CGpar,GTW,LeWin,Win}, in
order to analyze equations which do not have complete (as opposed
to asymptotic) symmetries -- as indeed for the equations
considered in this note. Our method is based on the abstract
approach developed in \cite{Gae}, based itself on ideas and
previous work by several authors \cite{Bar,BK,CE,Gold}.
\medskip

3.) The main body of this work was concerned -- and the method
described here can strictly speaking deal only -- with scale
invariance (at infinity or near a travelling front); this is
surely not the most general kind of (asymptotic or exact)
invariance, but definitely one of physical relevance. We trust
that suitable generalizations of our approach can also deal with
more general asymptotic invariance properties.
\medskip

4.) Here we focused on a restricted class of anomalous diffusion
and reaction-diffusion equations, which one of us had previously
studied numerically \cite{MVV2}, and on the equation for the
marginal Wigner distribution in an optical lattice, whose
stationary solutions are known to have certain asymptotic scaling
properties \cite{Abe,Lut}. It is however quite clear (also due to
the geometrical nature of the ideas at the basis of our
constructions) that the method developed and applied in this paper
does apply to quite general differential equations, and is
potentially capable to provide a sound explanation -- or
prediction -- of the asymptotic invariance of their solutions.


\medskip



\begin{thebibliography}{49}

\bibitem{Abe} S. Abe, ``Dilatation symmetry of the Fokker-Planck equation and anomalous diffusion'', {\it Phys. Rev. E} {\bf 69} (2004), 016102

\bibitem{AFT} I. Anderson, M. Fels and Ch. Torre, ``Group invariant solutions without transversality'', {\it Comm. Math. Phys.} {\bf 212} (2000), 653-686

\bibitem{BGI} V.A. Baikov, R.K. Gazizov and N.Kh. Ibragimov,  Approximate symmetries, {\it (Russian) Mat. Sb. (N.S.)}, {\bf  136(178)} (1988) 435-450; translation in {\it Math. USSR-Sb.} {\bf  64} (1989) 427-441

\bibitem{Bar} G.I. Barenblatt, {\it Scaling, self-similarity, and intermediate asymptotics}, Cambridge University Press, Cambridge 1996

\bibitem{Bar2} G.I. Barenblatt, {\it Scaling}, Cambridge University Press, Cambridge 2003

\bibitem{BlC} G.W. Bluman and J.D. Cole, {\it Similarity methods for differential equations}, Springer, Berlin 1974; G.W. Bluman and S. Kumei, {\it Symmetries of differential equations}, Springer, Berlin 1989

\bibitem{BK} J. Bricmont and A. Kupiainen, ``Renormalization group and the Ginzburg-Landau equation'', {\it Comm. Math. Phys.} {\bf 150} (1992), 193-208

\bibitem{CDW} J.F. Carinena, M. Del Olmo and P. Winternitz, ``On the relation between weak and strong invariance of differential equations", {\it Lett. Math. Phys.} {\bf 29} (1993), 151-163

\bibitem{CLV} M. Cencini, C. Lopez and D. Vergni, ``Reaction-diffusion systems: front propagation and spatial structures'', in: {\it The
Kolmogorov Legacy in Physics} (Lect. Notes Phys. 636), A. Vulpiani
and R. Livi eds., Springer, Berlin 2003

\bibitem{CiK} G. Cicogna, ``Weak symmetries and adapted
variables for differential equations'', {\it Int. J. Geom. Meth.
Mod. Phys.} {\bf 1} (2004), 23-31

\bibitem{CG} G. Cicogna and G. Gaeta, {\it Symmetry and perturbation theory in nonlinear dynamics} (LNPm57), Springer, Berlin 1999

\bibitem{CGpar} G. Cicogna and G. Gaeta, ``Partial Lie-point symmetries of differential equations'', {\it J. Phys. A} {\bf 34} (2001), 491-512

\bibitem{ClK} P.A. Clarkson and M.D. Kruskal,  New similarity reductions of the Boussinesq equation, {\it J. Math. Phys.} {\bf 30} (1989), 2201-2213

\bibitem{CE} P. Collet and J.P. Eckmann, {\it Instabilities and fronts in extended systems}, Princeton University Press, Princeton 1990

\bibitem{CrH} M.C. Cross and P.C. Hohenberg, ``Pattern formation outside equilibrium'', {\it Rev. Mod. Phys.} {\bf 65} (1993), 851-1112

\bibitem{EbS} U. Ebert and W. van Sarloos, ``Front propagation into unstable states: universal algebraic convergence towards uniformly translating pulled fronts'', {\it Physica D} {\bf 146} (2000), 1-99

\bibitem{Fis} R.A. Fisher, ``The Wave of Advance of Advantageous Genes'', {\it Ann. Eugenics} {\bf 7} (1937), 355-369

\bibitem{Fus1} V.I. Fushchich, N.I. Serov and V.I. Chopik,
Conditional invariance and nonlinear heat equations,  (Russian,  English summary) {\it Dokl. Akad. Nauk Ukrain. SSR Ser. A}  {\bf 86} (1988), 17-21; V.I. Fushchich, W.M. Shtelen and S.L. Slavutsky,
Reduction and exact solutions of the Navier-Stokes equations,  {\it  J. Phys. A: Math. Gen.}  {\bf 24} (1991), 971-984

\bibitem{Fus2} W.I. Fushchich and W.M. Shtelen,  On approximate symmetry and approximate solutions of the non-linear wave equation with a small parameter, {\it J. Phys. A: Math. Gen.} {\bf  22} (1989), L887-L890

\bibitem{Gae} G. Gaeta, ``Asymptotic symmetries and asymptotically symmetric solutions of partial differential equations'', {\it J. Phys. A} {\bf 27} (1994), 437-451

\bibitem{GaeK} G. Gaeta, {\it Nonlinear symmetries and nonlinear equations}, Kluwer, Dordrecht 1994

\bibitem{Gio} A. Giorgilli, ``Rigorous results on the power expansions for the integrals of a Hamiltonian system near an elliptic equilibrium point, {\it Ann. I.H.P. (Phys. Th\'eor.} {\bf 48} (1988), 423-439

\bibitem{Gold} N. Goldenfeld, O. Martin, Y. Oono and F. Liu, ``Anomalous diffusion and the renormalization group in a non-linear diffusion process'', {\it Phys. Rev. Lett.} {\bf 65} (1990), 1361-1364

\bibitem{GTW} A. M. Grundland, P. Tempesta and P. Winternitz,
``Weak Transversality and Partially Invariant Solutions'', {\it J.
Math. Phys.} {\bf 44} (2003), 2704-2722

\bibitem{Ibr} N.H. Ibragimov, ``Group analysis of ordinary differential equations and the invariance principle in Mathematical Physics'', {\it Russ. Math. Surv.} {\bf 47} (1992), 89-156

\bibitem{KPP} A.N. Kolmogorov, L.G. Petrovskii and N.S. Piskunov,
``Etude de l'equation de la diffusion avec croissance de la
matiere et son application a un probleme biologique'', {\it Bull.
Moscow Univ. Math. Mech.} {\bf 1} (1937), 1-25

\bibitem{Kra} I. S. Krasilshick and A.M. Vinogradov, {\it Symmetries and conservation laws for differential equations of mathematical physics}, AMS, Providence 1999

\bibitem{LeWin} D. Levi and P. Winternitz, ``Non-classical symmetry reduction: example of the Boussinesq equation'', {\it J. Phys. A} {\bf 22} (1989), 2915-2924

\bibitem{Lut} E. Lutz, ``Anomalous diffusion and Tsallis statistics in an optical lattice'', {\it Phys. Rev. A} {\bf 67} (2003), 051402(R)

\bibitem{MVV1} R. Mancinelli, D. Vergni and A. Vulpiani, ``Superfast front propagation in reactive systems with non-Gaussian diffusion'', {\it Europhys. Lett.} {\bf 60} (2002), 532-538

\bibitem{MVV2} R. Mancinelli, D. Vergni and A. Vulpiani, ``Front propagation in reactive systems with anomalous diffusion'', {\it Physica D} {\bf 185} (2003), 175-195

\bibitem{MeK1} R. Metzler and J. Klafter, ``The random walk's guide to anomalous diffusion: a fractional dynamics approach'', {\it Phys. Rep.} {\bf 339} (2000), 1-77

\bibitem{MeK2} R. Metzler and J. Klafter, ``The restaurant at the
end of the random walk: recent developements in the description of
anomalous transport by fractional dynamics'', {\it J. Phys. A}
{\bf 37} (2004), R161-R208

\bibitem{Mur} J.D. Murray, {\it Mathematical Biology}, Springer, Berlin 1993

\bibitem{Nak} M. Nakahara, {\it Geometry, Topology and Physics},
IOP, Bristol 1990

\bibitem{NaS} C. Nash and S. Sen, {\it Topology and Geometry for
physicists}, Academic Press, London 1983

\bibitem{Olv1} P.J. Olver, {\it Application of Lie groups to differential equations}, Springer, Berlin 1986

\bibitem{OlR} P.J. Olver and Ph. Rosenau, ``The construction of special solutions to partial differential equations'', {\it Phys. Lett. A} {\bf 114} (1986), 107-112;  ``Group invariant solutions of differential equations'', {\it SIAM J. Appl. Math.} \textbf{47} (1987), 263-278

\bibitem{OlV} P.J. Olver and E.M. Vorob'ev, ``Nonclassical and conditional symmetries'',
Chapter XI in {\it CRC Handbook of Lie group Analysis, vol.3}

\bibitem{Ovs} L.V. Ovsjannikov, {\it Group analysis of differential
equations}, Academic Press, New York 1982

\bibitem{PuS} E. Pucci and G. Saccomandi, ``On the weak symmetry group of partial differential equations'', {\it J. Math. Anal. Appl.} {\bf 163} (1992), 588-598

\bibitem{Ste} H. Stephani, {\it Differential equations.
Their solution using symmetries}, Cambridge University Press 1989

\bibitem{Tsa} C. Tsallis, ``Possible generalization of Boltzmann-Gibbs statistics'',
{\it J. Stat. Phys.} {\bf 52} (1988), 479-487

\bibitem{Wal} S. Walcher, ``Symmetries of ordinary differential equations'', {\it Nova J. Alg. Geom.} {\bf 2} (1993), 245-275; ``Orbital symmetries of first order ODEs'', in: {\it Symmetry and
Perturbation Theory (SPT98)}, A. Degasperis and G. Gaeta eds.,
World Scientific, Singapore 1999; M.K. Kinyon and S. Walcher, ``On
ordinary differential equations admitting a finite linear group of
symmetries'', {\it J. Math. Anal. Appl.} {\bf 216} (1997), 180-196

\bibitem{Win} P. Winternitz, ``Lie groups and solutions of nonlinear PDEs'', in {\it Integrable systems, quantum groups, and quantum field theory} (NATO ASI 9009), L.A. Ibort and M.A. Rodriguez eds., Kluwer,
Dordrecht 1993



\end{thebibliography}
\end{document}